\pdfoutput=1
\documentclass[sigconf,nonacm,balance=false]{acmart}

\setcopyright{none}
\renewcommand\footnotetextcopyrightpermission[1]{}
\usepackage{amsmath,amsfonts, amsthm}
\usepackage{algorithmic}
\usepackage{graphicx}
\usepackage{textcomp}
\usepackage{xcolor}
\usepackage{acronym}
\usepackage{subfig}
\usepackage{balance}
\usepackage{booktabs}
\usepackage{tabularx}
\usepackage{tcolorbox}
\usepackage{hyperref}

\newtheoremstyle{findingstyle}
    {0pt}                    %
    {0pt}                    %
    {\normalfont}            %
    {0pt}                    %
    {\bfseries}              %
    {. }                     %
    {0.0em}                  %
    {}                       %
\theoremstyle{findingstyle}

\newtheorem{findinginner}{Finding}[section]
\newenvironment{finding}
{\begin{tcolorbox}[
    colback=gray!10,              %
    colframe=gray,                %
    arc=5pt,                      %
    boxrule=1pt,                  %
    left=0pt,                     %
    right=0pt,                    %
    top=0pt,                      %
    bottom=0pt,                   %
    left skip=0pt,                %
    right skip=0pt                %
]%
\begin{findinginner}}
{\end{findinginner}%
\end{tcolorbox}}

\usepackage{todonotes}

\begin{document}

\acrodef{RFF}{Radio Frequency Fingerprinting}
\acrodef{RFFI}{Radio Frequency Fingerprint Identification}
\acrodef{RF}{Radio Frequency}
\acrodef{IQ}{In-Phase and Quadrature}
\acrodef{AE}{Autoencoder}
\acrodef{PSD}{Power Spectral Density}
\acrodef{CFO}{Carrier Frequency Offset}
\acrodef{DAC}{Digital-to-Analog Converter}
\acrodef{LoS}{Line-of-Sight}
\acrodef{LS}{Least-Squares}
\acrodef{RLS}{Recursive Least-Squares}
\acrodef{CNN}{Convolutional Neural Network}
\acrodef{ID}{In-Distribution}
\acrodef{OOD}{Out-of-Distribution}
\acrodef{MSE}{Mean Squared Error}
\acrodef{AUC}{Area Under the Receiver Operating Characteristic Curve}
\acrodef{FC}{Fully Connected}
\acrodef{BN}{Batch Normalization}
\acrodef{ReLU}{Rectified Linear Unit}
\acrodef{GPU}{Graphics Processing Unit}
\acrodef{SDR}{Software-Defined Radio}
\acrodef{BPSK}{Binary-Phase Shift Keying}
\acrodef{QPSK}{Quadrature-Phase Shift Keying}
\acrodef{DL}{Deep Learning}
\acrodef{DNN}{Deep Neural Networks}
\acrodef{ROC}{Receiver Operating Characteristic}

\title{Cross-Modulation Radio Frequency Fingerprinting}

\author{Gabriele Oligeri}
\affiliation{%
  \institution{College of Science and Engineering, Hamad Bin Khalifa University}
  \city{Doha}
  \country{Qatar}}
\email{goligeri@hbku.edu.qa}

\author{Ingrid Huso}
\affiliation{%
  \institution{College of Science and Engineering, Hamad Bin Khalifa University}
  \city{Doha}
  \country{Qatar}}
\email{ihuso@hbku.edu.qa}

\author{Savio Sciancalepore}
\affiliation{%
  \institution{Eindhoven University of Technology}
  \city{Eindhoven}
  \country{Netherlands}}
\email{s.sciancalepore@tue.nl}

\begin{abstract}
Radio Frequency Fingerprint Identification (RFFI) techniques exploit hardware-induced impairments embedded in transmitted signals to identify wireless transmitters. Existing RFFI studies typically assume that training and deployment data share the same physical-layer modulation. However, modern communication systems (e.g., WiFi, 5G) let devices dynamically adapt modulation parameters according to channel conditions, possibly causing the modulation observed during operation to differ from that used during model training. To our knowledge, this {\em cross-modulation RFFI} problem has not been systematically studied in the literature, especially in an open-set setting where only the model of a single authorized transmitter is available to the RFFI system.\\
In this paper, we systematically investigate cross-modulation RFFI in an open-set setting.
Using a dataset comprising twelve nominally identical software-defined radios, six modulation orders from 4-QAM to 256-QAM, and both wired and wireless line-of-sight links, we compare RFFI strategies operating on raw IQ samples collected at the RF front-end with image-based approaches leveraging symbols extracted after demodulation.
We show that transmitter-specific fingerprints largely persist in the received symbols across modulation changes, whereas they are significantly less stable in raw IQ samples. Building on this observation, we introduce a new image-based RFFI approach based on a pool of per-symbol models, enabling device identification even when the target modulation has not been observed during training. Our experimental results show a median area under the ROC curve of 0.975, compared with 0.928 achieved by a single-model approach and 0.816 achieved using raw IQ samples. These performance gains generalize to over-the-air deployments, even when training on cable-collected data.
\end{abstract}

\begin{CCSXML}
<ccs2012>
   <concept>
       <concept_id>10002978.10003014.10003017</concept_id>
       <concept_desc>Security and privacy~Mobile and wireless security</concept_desc>
       <concept_significance>500</concept_significance>
       </concept>
 </ccs2012>
\end{CCSXML}

\ccsdesc[500]{Security and privacy~Mobile and wireless security}

\keywords{Physical-layer security, open-set device identification, wireless security, radio measurements.}

\maketitle

\section{Introduction}
\label{sec:introduction}
\ac{RFFI} techniques exploit hardware-induced distortions left by the analog components of a radio transmitter to identify devices from their emitted signals~\cite{brik2008paradis}. Because it requires neither cryptographic material nor modifications to the transmitter, RFFI is regarded as a promising mechanism to authenticate legacy and resource-constrained devices and to complement higher-layer security mechanisms~\cite{jagannath2022survey}. Over the last years, \ac{DL} has become the reference tool for \ac{RFFI}, since \ac{DNN} can learn the fingerprint directly from the received signal, without hand-crafted features~\cite{jagannath2022survey}. At the same time, several works have shown that the learned fingerprint is fragile: a model trained under given conditions may fail when such conditions change, e.g., due to the wireless channel~\cite{alshawabka2020exposing}, the time elapsed since the training~\cite{alhazbi2023dat}, or radio reloads~\cite{oligeri2026wisec}. Thus, the robustness of the fingerprint to the conditions that change between training and testing is one of the main obstacles to the deployment of \ac{RFFI} in real systems. In this context, the modulation is possibly one of such conditions, but it has received little attention so far. Many modern wireless technologies adapt the modulation order to the quality of the link~\cite{goldsmith1997variable,80211n_adaptive,5g_link_adapt}, so the same device may change modulation over time, e.g., from 4-QAM to 256-QAM and back, depending on the channel conditions. Thus, a receiver willing to identify the transmitter via \ac{RFFI} should recognize it even when it transmits with another modulation scheme. We refer to this problem as \emph{Cross-modulation \ac{RFFI}}. At the same time, this problem is challenging, as the device fingerprint is a small perturbation superimposed on a signal whose statistical properties change significantly across modulation orders. As a consequence, models trained on one modulation may fail to generalize to another, while collecting training data for all possible modulation orders is often impractical, e.g., due to their occurrence only in specific link conditions which may be challenging to reproduce. As discussed in Sect.~\ref{sec:related_work}, the very few works considering a change of modulation between training and testing leave various issues unaddressed. First, they focus almost exclusively on closed-set identification, whereas real-world \ac{RFFI} systems must also reject previously unseen devices (i.e., work in open-set settings). Second, most proposed solutions require either samples of the target modulation or several modulation orders during training. Third, they operate on raw received samples, leaving open the question of whether transmitter fingerprints are better preserved across modulations before or after demodulation. Finally, all such studies consider modulation orders up to 64-QAM, and the feasibility of cross-modulation RFFI for high-order modulations is not investigated. Consequently, it remains unclear whether open-set cross-modulation RFFI is feasible at all and, if so, which signal representation should be used. Therefore, this paper addresses two main questions: (i) can a transmitter be identified via \ac{RFFI} when training and testing on different modulation orders in an open-set setting, and (ii) which signal representation best preserves the fingerprint across such changes?

{\bf Contribution.} To the best of our knowledge, this paper provides the first systematic experimental analysis of cross-modulation \ac{RFFI} in an open-set setting. We collect real-world radio measurements from twelve nominally identical \acp{SDR} transmitting with six modulation orders, from 4-QAM to 256-QAM, over both a cable and a wireless \ac{LoS} link. Using such data, we compare the performance of two \ac{RFFI} approaches: an \ac{AE} working on the \ac{PSD} of the raw \ac{IQ} samples and an \ac{AE} working on images generated from the demodulated symbols at the end of the RF reception chain. We show that the fingerprint observed on the symbols is largely preserved when the modulation changes, as opposed to the one observed on the raw \ac{IQ} samples. Building on such considerations, we propose a new approach based on a pool of per-symbol models, that handles a modulation never observed at training time without re-training, by processing each of its symbols through the models of the closest training constellation points, achieving a median \ac{AUC} of 0.975. Finally, we show that these results hold over the air, even when training on cable data and testing on wireless link data.

{\bf Paper organization.} Section~\ref{sec:related_work} reviews the related work, Sects.~\ref{sec:scenario} and~\ref{sec:setup} introduce the scenario and the experimental setup; Sect.~\ref{sec:methodology} presents the \ac{RFFI} methodology; Sect.~\ref{sec:results} reports the results on the cable, over the air, and across scenarios, and finally, Sect.~\ref{sec:conclusions} concludes the paper.

\section{Related Work}
\label{sec:related_work}
The goal of \ac{RFFI} is to identify a transmitter from the imperfections of its analog components, and this problem has been widely studied since the pioneering work of Brik et al.~\cite{brik2008paradis}. The vast majority of the available solutions train and test the model on signals with the same waveform, and investigate the robustness of the fingerprint to other sources of variability, such as the wireless channel~\cite{alshawabka2020exposing} and the time elapsed between the collection of the training dataset and the one of the testing dataset~\cite{alhazbi2023dat}. For example, Al-Shawabka et al.~\cite{alshawabka2020exposing} demonstrated that the wireless channel severely degrades the identification of devices with identical \ac{RF} circuitry, while equalizing the \ac{IQ} samples partially recovers the accuracy. The modulation of the waveform is yet another source of variability for \ac{RFFI}, since the information-bearing component of the signal is much stronger than the fingerprint. However, the literature has paid considerably less attention to the impact of modulation on \ac{RFFI}. A first group of works acknowledges that the modulation affects the fingerprint, without however investigating the impact of a change in the modulation between training and testing. For example, Tyler et al.~\cite{tyler2022intentional} removed the known structure of the IEEE 802.11a preamble from the signals collected from 32 commercial Wi-Fi adapters, and showed that the inclusion of such a known structure before the payload hinders the learning of the fingerprint. Wong et al.~\cite{wong2018iqimbalance} estimated the \ac{IQ} imbalance associated to each transmitter, which does not depend on the modulation, using a \ac{CNN} trained on simulated signals with several PSK and QAM orders. Agadakos et al.~\cite{agadakos2020chameleons} trained a single complex-valued network on the raw \ac{IQ} samples of Wi-Fi and ADS-B devices, with no knowledge of the protocol. Finally, Huang et al.~\cite{huang2024dtp} generated images from the symbols generated by five \acp{SDR} with \ac{BPSK} and 4-, 16-, and 64-QAM, but trained and tested only one model per modulation. Only a few works consider the same problem as this paper, namely variable-modulation or cross-modulation specific emitter identification, i.e., train a model with one modulation and test it with another one. Most of them treat the change of modulation as a domain shift. For example, Zhang et al.~\cite{zhang2023variable} fed both the received signal and its demodulated and reconstructed version to a domain-adversarial network, whereas Yin et al.~\cite{yin2023fewshot} fine-tuned a model trained on 16-PSK signals with a few labeled \ac{BPSK} signals. Similarly, Zhang et al.~\cite{zhang2025mdd} aligned the features of the two modulations through the margin disparity discrepancy, and He and Zhang~\cite{he2025crossmod} adapted a pre-trained model without accessing its training data. Nevertheless, all these domain adaptation approaches require signals with the new modulation during training, either labeled~\cite{yin2023fewshot} or unlabeled~\cite{zhang2023variable,zhang2025mdd,he2025crossmod}, as well as a new training stage whenever the modulation changes.
To overcome this limitation, some recent approaches rely on domain generalization, i.e.,  train on several modulations and test on unseen ones. This can be achieved either by disentangling the modulation-specific and device-specific features through a variational autoencoder~\cite{shi2026diva}, or by means of a similarity-aware mixup of the training modulations~\cite{wang2026diame}, or by combining the reconstructed signal with a lightweight calibration on the test signals~\cite{wang2026rsfe}. In this context, the works closest to ours are those of Chen and Xu~\cite{chen2024agnostic}, who modeled, for each emitter, the mapping from the reconstructed ideal signal to the received one and identified the emitter through a hypothesis test, thus extending models trained on \ac{BPSK} or \ac{QPSK} up to 32-QAM. Building on this idea, they further extended their work by decomposing the received signal into a fingerprint and a modulation component, by means of a generative model~\cite{chen2025disentangled}. However, both works consider closed-set \ac{RFFI} and synthetic data.

\begin{table}[t]
\caption{Overview of scientific papers considering \ac{RFFI} with waveforms characterized by various modulations. n.s. indicates that the paper did not specify the information.}
\label{tab:related}
\centering
\footnotesize
\setlength{\tabcolsep}{3pt}
\resizebox{\columnwidth}{!}{%
\begin{tabular}{|l|l|c|c|c|c|c|}
\hline
\textbf{Ref.} & \textbf{Data} & \textbf{Tx} & \textbf{Max. QAM} & \textbf{Channel} & \textbf{Adapt.-free} & \textbf{Open set} \\ \hline
\cite{zhang2023variable} & Simulated & 7 & 64 & Sim.\ fading & No & No \\
\cite{yin2023fewshot} & Real & 6 & PSK only & n.s. & No & No \\
\cite{chen2024agnostic} & Simulated & 7 & 32 & Sim.\ AWGN & Yes & No \\
\cite{zhang2025mdd} & Real & 7 & 64 & n.s. & No & No \\
\cite{he2025crossmod} & Simulated & 10 & 16 & Sim.\ fading & No & No \\
\cite{chen2025disentangled} & Simulated & 6 & 64 & Sim.\ AWGN & Yes & No \\
\cite{shi2026diva} & Simulated, real & 5 & 64 & Sim.\ AWGN & Yes & No \\
\cite{wang2026diame} & Simulated & 5 & 16 & Sim.\ AWGN & Yes & No \\
\cite{wang2026rsfe} & Real & 6 & PSK only & LoS + sim.\ fading & Partially & No \\ \hline
\textbf{This work} & Real & 12 & 256 & Wired, wireless LoS & Yes & Yes \\ \hline
\end{tabular}}
\end{table}

{\bf Research Gap.} As summarized in Table~\ref{tab:related}, our contribution differs from the literature in four aspects. First, all the literature on \ac{RFFI} considering various modulation orders addresses closed-set \ac{RFFI}, i.e., the classification of a signal among a set of known transmitters. In contrast, we address open-set \ac{RFFI}, in which the receiver trains only on one authorized transmitter and has to identify and reject any other device without any prior knowledge about them. Notably, to our knowledge, open-set \ac{RFFI}, even by means of \acp{AE}, has so far been investigated only considering a single waveform only~\cite{hanna2021openset,oligeri2023pastai,smailes2023watch}. Second, only the solutions in~\cite{chen2024agnostic,chen2025disentangled} are adaptation-free, i.e., they rely on a single modulation per emitter during training, while the others require either signals with the new modulation~\cite{zhang2023variable,yin2023fewshot,zhang2025mdd,he2025crossmod} or several training modulations~\cite{shi2026diva,wang2026diame,wang2026rsfe}. However, the approach in~\cite{zhang2023variable} discards the received samples that fall outside the training constellation, which become the majority as the order increases, and both the adaptation-free approaches rely on a hypothesis test among emitters that must all be known at training time. Conversely, the single-model approach proposed in this paper performs \ac{RFFI} using all symbols of the previously unseen modulation, whereas the pool of per-symbol models associates each constellation point of the new modulation with the closest constellation points observed during training.. Third, most of the works considering cross-modulation \ac{RFFI} rely on simulated data~\cite{zhang2023variable,he2025crossmod,shi2026diva,wang2026diame,chen2024agnostic,chen2025disentangled}, generated by adding noise or simulated fading to synthetic data, and none of them considers modulation orders greater than 64-QAM. In contrast, we consider twelve radios of the same model, a wired link and an indoor line-of-sight wireless link, and all the pairs of modulations from 4-QAM to 256-QAM, which reflect the order changes of adaptive modulation schemes. Finally, while the existing solutions process the samples of the received signal and adopt demodulation only to reconstruct a reference signal~\cite{zhang2023variable,he2025crossmod,wang2026rsfe,chen2024agnostic}, we systematically compare \ac{IQ} samples and symbols as input of the \ac{RFFI} model, and investigate which approach maximizes the effectiveness of cross-modulation \ac{RFFI}.

\section{Scenario}
\label{sec:scenario}
\begin{figure}[t]
 \centering
 \includegraphics[width=\columnwidth]{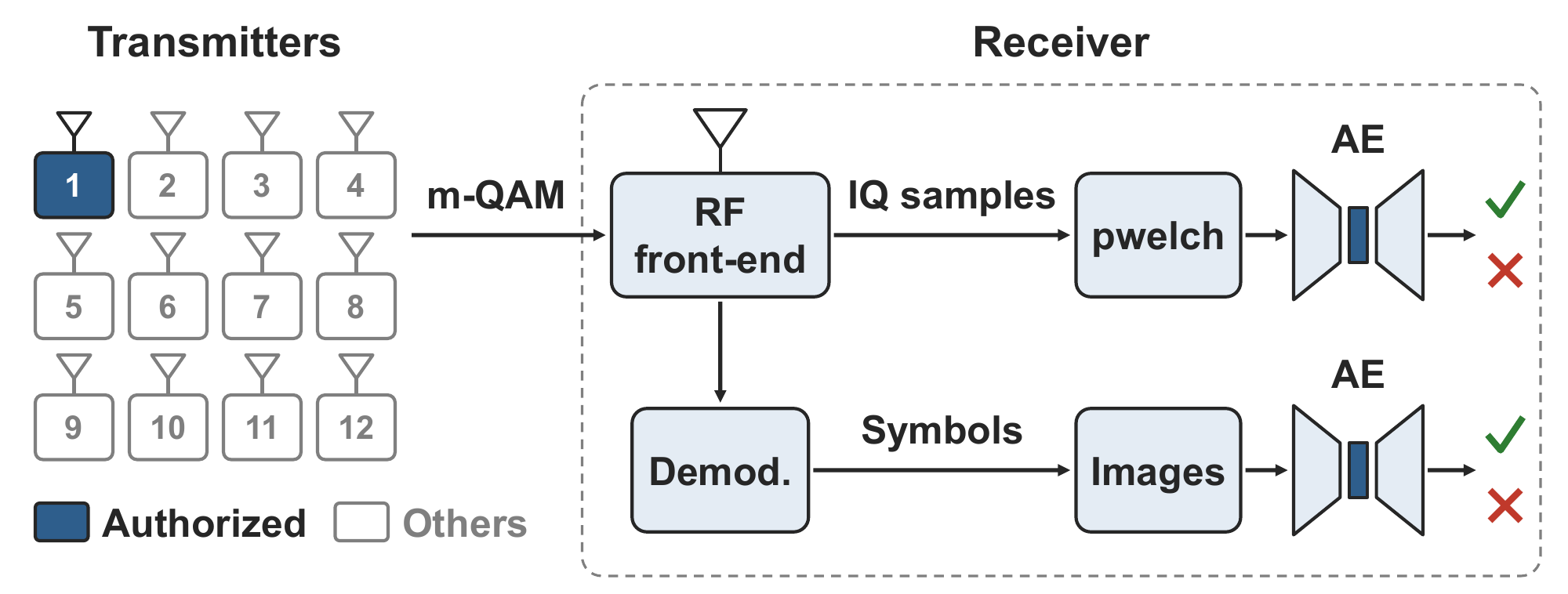}
 \caption{Reference scenario: the receiver identifies one authorized transmitter out of a pool of twelve possible ones using \ac{RFFI}, either directly from the IQ samples delivered by the RF front-end (upper branch) or by analyzing the symbols at the output of the demodulator (lower branch). In both cases, the pre-processed data are fed to an autoencoder, which can accept or reject the transmitter.
 }
 \Description{Block diagram: twelve transmitters, one of them marked as authorized, and a receiver with two branches from the RF front-end to an autoencoder, one through a pwelch block and one through a demodulator and an images block.}
 \label{fig:scenario}
\end{figure}
In our reference scenario, we consider a pool of transmitters and one receiver, as depicted in Fig.~\ref{fig:scenario}. The transmitters are devices of the same model, running the same software with the same communication parameters, so that their signals cannot be distinguished solely by analyzing the transmitted content. The receiver aims to {\em identify} the transmitter at the physical layer, i.e., verify that the received signal has been generated by a given authorized device of the pool and not by any of the others, by relying only on the imperfections that the hardware of each transmitter leaves on the signal. To this end, the receiver observes the authorized transmitter during a training phase and builds a model of its fingerprint. The remaining transmitters are not available to the receiver at training time, and they must be rejected when they are later observed. Moreover, we consider that the transmitter may change its modulation order over time, as it happens with adaptive modulation schemes~\cite{svensson2007adaptive,5g_link_adapt}. Thus, the receiver is required to identify the authorized transmitter even when the modulation order used at testing time differs from the one observed during training. We refer to this scenario as {\em cross-modulation \ac{RFFI}}. In line with the literature on \ac{RFFI}, the receiver can perform physical-layer device identification at different stages of its processing chain. These stages correspond to the two branches shown in Fig.~\ref{fig:scenario}, which provide different types of data: (i) the \ac{IQ} samples available just outside the \ac{RF} front-end, before any processing, and (ii) the demodulated symbols, as available at the output of the demodulator. The two configurations expose different information on the transmitter and place different requirements on the receiver, and each of them comes with a trade-off that potentially affects the performance of \ac{RFFI}. In both cases, the receiver pre-processes the collected data (\emph{pwelch} and \emph{Images} blocks in Fig.~\ref{fig:scenario}, see Sect.~\ref{sec:preprocessing}) before feeding them to an \ac{AE}, which can either accept or reject the transmitter.

{\bf Raw IQ-samples.} In the first configuration, the receiver processes the raw \ac{IQ} samples as they are detected and provided by the \ac{RF} front-end. To generate such samples, the receiver requires neither synchronization nor demodulation of the signal. Thus, the receiver can identify the transmitter from a short portion of the signal taken at any time, without knowledge of the frame structure or of the modulation, and even when the signal cannot be decoded. At the same time, the receiver cannot apply any further processing to the signal to improve its quality. Together with the fingerprint, the samples carry the transmitted message content, the \ac{CFO} and the effect of the channel, and the receiver has to isolate the contribution of the fingerprint from all of them. In addition, the amplitude distribution of the samples changes with the modulation order. Such changes may alter the way the impairments of the transmitter appear in the signal, and thus hinder the transfer of a model from one modulation to another.

{\bf Symbols.} In the second configuration, the receiver first recovers the symbols associated with the received \ac{IQ} samples, and then it performs transmitter identification at the end of the decoding chain. Through synchronization and equalization, the receiver can remove the \ac{CFO} and most of the effect of the channel. Moreover, the receiver can use the knowledge of the transmitted symbols to separate the content from the distortion, to observe the impairments of the transmitter directly as the displacement of the received symbols from the ideal constellation points. However, these advantages come at the cost of relying on the entire reception chain, as the receiver must detect and decode the frames, have prior knowledge of the modulation scheme, and collect a sufficient number of symbols before making a decision.
Furthermore, the equalizer used for preamble detection may remove part of the distortions introduced by the transmitter together with that of the channel, and the constellation itself changes with the modulation order. Thus, a fingerprint learned on the points of one constellation does not necessarily apply to the points of another.

Therefore, it is not straightforward to determine a priori which of the two configurations provides the best identification performance, and to what extent this performance is preserved across modulation changes. The remainder of the paper addresses these questions by applying the same methodology to both types of data.

\section{Experimental Setup and Measurement Settings}
\label{sec:setup}

To match the scenario described in Sect.~\ref{sec:scenario} as closely as possible, we deployed an experimental testbed and collected real-world data for further analysis.

{\bf Hardware.} We deployed twelve USRP B200mini-i \acp{SDR} acting as transmitters and one USRP X410 acting as receiver. The transmitters are nominally identical, so any difference among their signals originates from the manufacturing tolerances of their \ac{RF} front-ends. The receiver observes each transmitter over two different links, with the same hardware and software configuration. In the first scenario, namely \emph{RF cable}, we connect each of the transmitters to the receiver through a coaxial cable and a 20~dB attenuator. Thus, in this scenario, the channel is flat and time-invariant, allowing us to isolate the hardware impairments of the transmitter. In the second scenario, namely \emph{wireless}, we remove the cable and the attenuator, and the radios communicate over the air through VERT900 omnidirectional antennas, one at each end of the link. We placed the transmitters and the receiver in the same office, 2~m apart, with a \ac{LoS} link between their antennas. We placed the transmitters one at a time at the same position, so that all devices experience the same propagation geometry. Moreover, one person is present in the office during all the measurements.

{\bf Software.} We control both ends of the link using MATLAB R2026a, through the Communications Toolbox Support Package for USRP Radio, with the parameters set as summarized in Table~\ref{tab:sw}. We designed the transmitted signal to meet two requirements of our analysis: (i) the modulation order was the only parameter varied across recordings of a given transmitter, and (ii) the transmitted symbols were known at the receiver. To this aim, we configured each transmitter to send a continuous stream of frames. Each frame consists of a 200-symbol preamble modulated via \ac{QPSK}, known to the receiver and scaled to the peak amplitude of the payload constellation, followed by a 500-symbol $m$-QAM payload, using Gray mapping and unit average power. We configured the first five payload symbols to encode the sequence number of the frame, while the remaining 495 payload symbols are pseudo-random symbols seeded with the same sequence number. Therefore, the receiver can regenerate the transmitted payload of every frame, which provides ground-truth symbol labels without a side channel. Moreover, we used a fixed digital back-off (waveform peak at 0.7 of the \ac{DAC} full scale) and a fixed transmit gain to keep the operating point of the transmit chain independent of $m$. We do not use automatic gain control at either end of the link.

\begin{table}[t]
\caption{Configurations common to the two scenarios.}
\label{tab:sw}
\centering
\footnotesize
\begin{tabular}{|l|l|}
\hline
Carrier frequency, sample rate & 900~MHz, 2~MS/s \\
Samples per symbol, symbol rate & 8, 250~kBd \\
Pulse shaping & RRC, roll-off 0.5, span 10 symbols \\
Frame (2.8~ms) & 200 QPSK + 500 $m$-QAM symbols \\
Modulation order $m$ & 4, 16, 32, 64, 128, 256 \\
Digital peak level & 0.7 of full scale \\ \hline
\end{tabular}
\end{table}

\begin{table}[t]
\caption{Parameters of the two scenarios.}
\label{tab:scenarios}
\centering
\footnotesize
\setlength{\tabcolsep}{4pt}
\begin{tabular}{|l|c|c|}
\hline
 & \textbf{RF cable} & \textbf{Wireless LoS} \\ \hline
Link & Coaxial cable, & Over the air, \\
 & 20~dB attenuator & office, LoS \\
Tx--Rx distance & -- & 2~m \\
Antennas & -- & VERT900 \\
Environment & -- & One person present \\
Tx gain setting (0--89.8~dB) & 60~dB & 60~dB \\
Rx gain setting (0--60~dB) & 35~dB & 35~dB \\
Recordings & 72 & 72 \\
Recording duration & 100~s & 140~s \\
Frames per recording & $3.57\times10^{4}$ & $4.99\times10^{4}$ \\
Payload symbols per recording & $1.79\times10^{7}$ & $2.49\times10^{7}$ \\
Average symbol error rate & $1.7\times10^{-6}$ & $2.9\times10^{-5}$ \\
Average error vector magnitude & 1.7\% & 1.7\% \\ \hline
\end{tabular}
\end{table}

{\bf Measurement Settings.} In each scenario, we recorded the raw \ac{IQ} samples available at the receiver, as received from every transmitter with all six modulation orders, one after the other and without altering the setup, for a total of $12\times6\times2=144$ recordings. Table~\ref{tab:scenarios} compares the two scenarios. We kept the transmit and receive gains the same, so that the two scenarios differ only in the link and in the duration of the recordings. Such gains represent the settings of the two front-ends, i.e., relative values in dB within the control range of the respective radio (0--89.8~dB for the transmitter and 0--60~dB for the receiver). We did not perform previous calibration of the power levels. The recordings in the \emph{RF cable} scenario last 100~s, so including $2\times10^{8}$ complex baseband samples and approx. $3.57\times10^{4}$ frames, while the recordings in the \emph{wireless} one last 140~s, including $2.8\times10^{8}$ samples and approx. $4.99\times10^{4}$ frames. We used a longer duration of the \emph{wireless} recordings as a safety margin: it compensates for the frames that might be lost over the air during the measurement campaign, so that each recording provides at least as many frames as in the \emph{RF cable} scenario. In practice, the two scenarios provide approx. the same number of frames per second (357), i.e., the losses turned out to be negligible. The link is reliable in both scenarios, with an average symbol error rate below $10^{-4}$ and the same average error vector magnitude. We configured the receiver to store only the raw baseband samples, while all the subsequent processing is performed offline, to minimize storage requirements.\\
In line with the scope of our analysis (see Sect.~\ref{sec:scenario}), we consider two types of data: i) the stored raw \ac{IQ} samples, and ii) the received symbols, which we obtained by demodulating the raw \ac{IQ} samples with a burst-mode receiver that relies only on the preamble of each frame. Specifically, after a coarse estimate of the \ac{CFO}, we configured the receiver to detect frames through the normalized cross-correlation with the pulse-shaped preamble, whose peak also provides the symbol timing. In the receiver chain, the receiver refines the \ac{CFO} estimate of each frame through a \ac{LS} fit of the preamble phase. Herein, the receiver trains a symbol-spaced, 7-tap \ac{RLS} equalizer on the preamble and operates in decision-directed mode over the payload, so that it compensates the residual carrier phase, the link gain, the fractional timing offset, and, in the wireless scenario, the channel dispersion. Finally, the receiver de-maps the equalized payload, and uses the decoded sequence number to regenerate the transmitted symbols, so that every received symbol is paired with the constellation point that was actually transmitted.

\section{RFFI Methodology}
\label{sec:methodology}

\subsection{Data Pre-processing}
\label{sec:preprocessing}
The objective of this work is to compare the performance of cross-modulation \ac{RFFI} when analyzing two different types of data, taken at the two ends of the receiver chain: (i) the raw \ac{IQ} samples delivered by the \ac{RF} front-end, and (ii) the symbols available at the output of the demodulator. Note that we cannot feed either type of data to an \ac{AE} as is. Both the time-domain \ac{IQ} samples and the sequence of received symbols are dominated by the transmitted payload, which is random and changes from frame to frame, whereas the impairments of the transmitter are a small perturbation superimposed on such a signal. An \ac{AE} trained on such data could learn to reconstruct any valid signal, irrespective of the device that generated it, and so its reconstruction error would carry no information on the transmitter. Therefore, in our methodology, we convert each type of data into a fixed-size representation that does not depend on the transmitted content, while retaining the distortion introduced by the transmitter: the \ac{PSD} for the \ac{IQ} samples, and images of the error vectors for the symbols. For both choices, we follow the approaches that the literature has consolidated for the respective type of data, as discussed in the following. At the same time, such choices differ in the knowledge required at the receiver: the former needs no synchronization or demodulation, whereas the latter relies on the complete receiver chain. In both cases, we used the first 70\% of each recording for training and the last 30\% for testing.

\begin{figure}[t]
 \centering
 \includegraphics[width=\columnwidth]{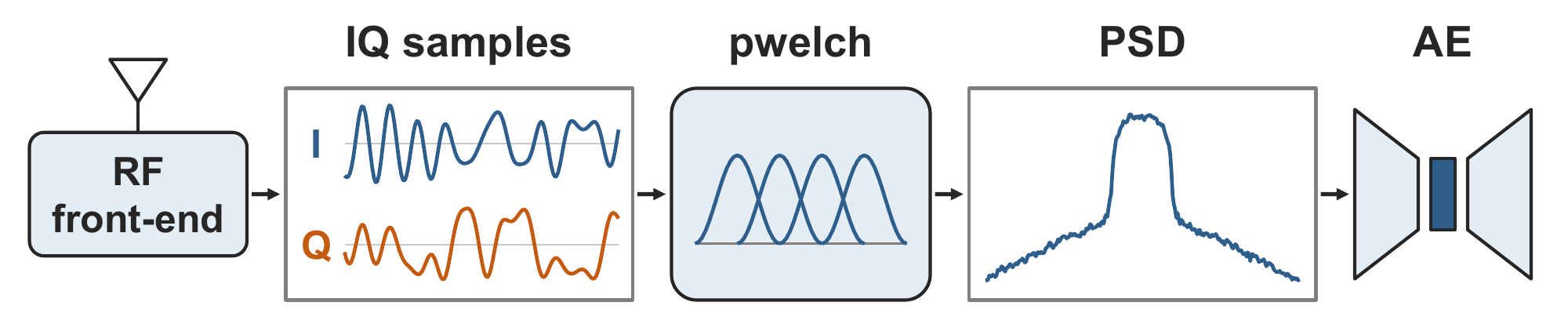}
 \caption{Pre-processing of the raw IQ samples: we convert a window of 4000 IQ samples delivered by the RF front-end, taken without synchronization or demodulation, through the Welch's method (\emph{pwelch}), into a PSD vector of 256 frequency bins, which is the input of the AE.}
 \Description{Block diagram: RF front-end, IQ samples, pwelch, PSD and AE, connected from left to right.}
 \label{fig:iq_pipeline}
\end{figure}

{\bf Raw IQ-samples.} For signals observed before demodulation (raw \ac{IQ}), we use as the input of the \ac{AE} the \ac{PSD} of the received signal, estimated with the Welch's method~\cite{welch1967}. Four arguments support this choice. First, the impairments of the transmitter leave their signature in the spectrum of the signal. The nonlinearity of the power amplifier generates spectral regrowth, i.e., a widening of the spectrum outside the signal band, which has been exploited to identify emitters by several seminal~\cite{liu2008nonlinear} and recent works~\cite{xie2025regrowth}. Similarly, works such as~\cite{elmaghbub2020widescan} have exploited the out-of-band distortion in addition to the in-band information to improve the classification of devices with the same hardware. In addition, Shi et al.~\cite{shi2026psd2san} recently derived how the \ac{IQ} imbalance, the oscillator distortion and the nonlinearity of the power amplifier shape the \ac{PSD}, and relied on the \ac{PSD} estimated with Welch's method to measure the similarity between the emitters of different domains. Second, spectral features are among the earliest and most established fingerprints used in the literature. Suski et al.~\cite{suski2008spectral} used the \ac{PSD} of the received signal for \ac{RFFI}, Kennedy et al.~\cite{kennedy2008steady} used the spectrum of the steady-state portion of the transmission, and Cai et al.~\cite{cai2022uav} used the power spectrum estimated with Welch's method among the fingerprint features adopted to identify unmanned aerial vehicles. Moreover, the survey in~\cite{jagannath2022survey} observes that the side lobes of the \ac{PSD} carry most of the information on the identity of the emitter. Third, the literature has proved that spectral representations are effective as the input of \ac{DNN}. For example, Shen et al.~\cite{shen2021lora} demonstrated that a time-frequency representation outperforms the direct use of the \ac{IQ} samples as input of a \ac{CNN}. Moreover, Scholl et al.~\cite{scholl2026dsp} recently combined a power spectrum obtained by averaging Hann-windowed segments with 50\% overlap, i.e., the estimator adopted in this work, with the \ac{IQ} samples to distinguish different implementations of a transmitter from the radiated signal~\cite{scholl2026dsp}. Fourth, contributions such as~\cite{saifaldawla2023ngso,saifaldawla2024genai} used this same processing chain to detect interfering transmitters in satellite links, while others used it to detect anomalous transmissions in the wireless spectrum~\cite{rajendran2019saife}.
The use of the Welch's method requires neither synchronization nor knowledge of the modulation. Moreover, by averaging the periodograms of several segments, the Welch's method reduces the dependence on the transmitted content, which changes from window to window, while retaining the spectral shape that carries the impairments. Thus, this choice also fits the first configuration discussed in Sect.~\ref{sec:scenario}. Taking into account all such considerations, in this paper, we adopt the \ac{PSD} as the representation of choice for a receiver that does not possess either the frame timing or a reference signal.
We show in Fig.~\ref{fig:iq_pipeline} the resulting processing chain, from the \ac{RF} front-end to the \ac{AE}. We divide the stream of \ac{IQ} samples delivered by the \ac{RF} front-end into windows of 4000 consecutive samples (duration 2~ms), each starting at an arbitrary position of the recording. We apply no frame synchronization, \ac{CFO} correction, filtering, or demodulation, so that a window contains an arbitrary portion of preamble and payload, with unknown symbol timing and carrier phase. As illustrated by the in-phase (I) and quadrature (Q) components in the \emph{IQ samples} block, we determine the waveform of a window from the transmitted symbols, and therefore such a waveform changes from window to window. We process each window through the \emph{pwelch} block, containing two parts: (i) generation of Hann-windowed segments of 256 samples with 50\% overlap, as sketched inside the block, and (ii) averaging of the segments of the periodograms. The output, shown in the \emph{PSD} block, is a vector of 256 frequency bins spanning the whole 2~MHz band, with a resolution of 7.8~kHz.
Note that, unlike the \ac{IQ} samples, the \ac{PSD} does not depend on the transmitted symbols. The \ac{PSD} averages out the contribution of the \ac{IQ} samples over the segments, while its shape retains the effect of the transmitter impairments: both the in-band portion and the out-of-band skirts visible in Fig.~\ref{fig:iq_pipeline} are part of the representation. Moreover, the \ac{PSD} is insensitive to the carrier phase and to the symbol timing, and the \ac{CFO} is negligible with respect to the bin width. The final two steps of our chain include two normalization operations. First, we express the \ac{PSD} in dB and normalize it to its average level within $\pm100$~kHz. This operation removes the received power component, so that only the spectral shape is retained and the classifier cannot exploit the link attenuation for \ac{RFFI}. Second, we standardize each frequency bin to zero mean and unit variance, using the statistics of the training windows of the transmitter observed during training, and we feed the resulting vector to the \ac{AE}. From each recording, we selected 2000 training and 500 test windows, evenly spaced in time within the respective partition, for a total of $1.8\times10^{5}$ windows per scenario.

{\bf Symbols.} For data observed after demodulation (symbols), the literature uses as the reference fingerprint the deviation of the received symbols from the ideal constellation points, as introduced in~\cite{brik2008paradis}, through a set of modulation-domain errors, while Peng et al. later combined constellation-based figures with \acp{CNN}~\cite{peng2020dctf}.
In particular, the literature has shown that the conversion of the demodulated samples of the \ac{IQ} plane into images, which are then processed by \ac{DNN}, is particularly effective and robust for satellite transmitters~\cite{oligeri2023pastai}, for the stability of the fingerprint over time~\cite{alhazbi2023dat}, against adversarial machine learning~\cite{papangelo2024adversarial}, and for the study of fingerprint obfuscation~\cite{irfan2025jamming,oligeri2026hideprint}. Many contributions, e.g., ~\cite{oligeri2023pastai,oligeri2026hideprint,sciancalepore2024jamming}, adopted \acp{AE} trained on such images for one-class detection. We adopt this methodology with one modification, required by our cross-modulation setting: instead of a single image of the whole constellation, we generate one image per constellation point (symbol), built from the error vectors on a grid that is common to all modulation orders. This choice allows us to generalize one-class image-based approaches to a setting where the transmitters could use several modulations, and not all of them may be known at training time (see Sect.~\ref{sec:symbols_method} for more details).

\begin{figure}[t]
 \centering
 \includegraphics[width=\columnwidth]{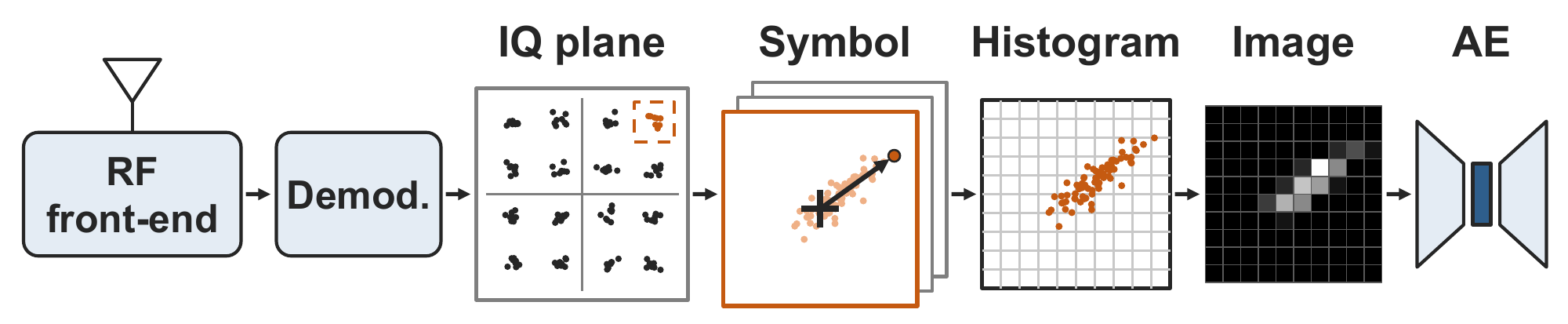}
 \caption{Pre-processing of the symbols: after demodulation (\emph{Demod.}), we extract the received symbols of each constellation point from the IQ plane (\emph{Symbol}), and we accumulate into a bivariate histogram their error vectors (the arrow shows one of the highlighted symbols) with respect to the ideal point (cross). We store such images as a $32\times32$ greyscale image, and we feed them to the AE.}
 \label{fig:symbol_pipeline}
\end{figure}

Figure~\ref{fig:symbol_pipeline} shows the resulting processing chain. Differently from the previous case, we take the data at the end of the decoding chain. We first process the output of the \ac{RF} front-end through the demodulator (\emph{Demod.} block), which performs the synchronization and the equalization described in Sect.~\ref{sec:setup}. Then, we normalize the received symbols of each recording to unit average power without removing the mean, so as to preserve the carrier leakage of the transmitter. The \emph{IQ plane} block in Fig.~\ref{fig:symbol_pipeline} shows the result for 16-QAM: the received symbols gather in a cloud around each constellation point. Then, we separate the clouds (\emph{Symbol} block): we associate each received symbol with the constellation point that was actually transmitted, as obtained from the regenerated payload rather than with the decision of the receiver, so that the cloud of a constellation point is not truncated at the decision boundaries.
For each received symbol, we compute the error vector, i.e., the difference between the received symbol and the ideal constellation point. In the \emph{Symbol} block of Fig.~\ref{fig:symbol_pipeline}, the cross represents the ideal point, the dots are the received symbols, and the arrow is the error vector of one of them, i.e., the dot with the black edge. As exemplified in the figure, in general, the cloud of the error vectors is neither centered on the ideal point nor circular: its displacement and its shape reflect the impairments of the transmitter and constitute its fingerprint. Since the error vectors are referred to the ideal point, they do not depend on the transmitted symbol. Finally, we convert the error vectors into images. We accumulate a block of 320 consecutive error vectors of the same class, i.e., of the same transmitter, modulation order, and constellation point, into a bivariate histogram (\emph{Histogram} block) of $32\times32$ bins covering the region $[-0.115,0.115]^2$ of the error plane. We store the histogram as an 8-bit greyscale image, whose pixel values are the bin counts (\emph{Image} block), and this block is the input of the \ac{AE}.

Recall from Sect.~\ref{sec:setup} that the payload symbols are equiprobable. However, since each modulation order $m$ includes different possible symbols, we require a different amount of time to collect all symbols of a constellation of the modulation order $m$. On average over $320m$ payload symbols, i.e., about $1.8m$~ms of signal with the frame structure of Table~\ref{tab:sw}, this time varies from about 7~ms for 4-QAM to about 0.46~s for 256-QAM. At the same time, the receiver collects the symbols of all the $m$ constellation points, i.e., one image every 1.8~ms on average. We kept the region and the bin width the same for every modulation order, and we do not rescale the bin counts: a pixel represents the same error vector in every image, which makes the images of different modulations directly comparable. For each class, we generate 140 training and 60 test images from blocks evenly spaced over the respective partition, so that the classes are balanced and span the whole recording. With 12 transmitters and 500 constellation points over the six modulation orders, each scenario provides 6000 classes and $1.2\times10^{6}$ images.

\subsection{Autoencoder-Based Open-Set Identification}
\label{sec:ae}
The \ac{PSD} vectors and the images generated as described in Sect.~\ref{sec:preprocessing} represent the inputs of two \acp{AE}, one for each type of data, which we use in the same way. We formulate transmitter identification as a one-class, open-set problem: for each authorized transmitter and modulation, we train an \ac{AE} only on the \ac{ID} samples of that transmitter, while the remaining eleven transmitters are \ac{OOD}, which the \ac{RFFI} system never observes during training. Since the \ac{AE} learns to reconstruct only the samples of the transmitter observed during training, we expect that the samples of any other device are reconstructed with a larger error. Therefore, at test time, the anomaly score of a sample is the \ac{MSE} between the sample and its reconstruction, and we attribute the sample to the authorized transmitter if the score is below a threshold. We base each decision on a single \ac{IQ} window or on a single image. We measure performance through the \ac{AUC} of the \ac{ROC} curve, which does not depend on the choice of the threshold. To assess cross-modulation generalization, we also evaluate a model trained with modulation order $m_A$ on the test samples of every other order $m_B$. We summarize in Table~\ref{tab:ae} the configuration of the two \acp{AE}. Note that we configure their architectures to match the structure of the respective inputs, as detailed in the following.
\begin{table}[t]
\caption{\ac{AE} network architectures.}
\label{tab:ae}
\centering
\footnotesize
\setlength{\tabcolsep}{4pt}
\begin{tabular}{|l|l|l|}
\hline
 & \textbf{Raw IQ-samples} & \textbf{Symbols} \\ \hline
Input & 256-bin PSD vector & $32\times32\times1$ image \\
Encoder & FC 64, ReLU; & 3 $\times$ [conv $3\times3$, BN, \\
 & FC 16, ReLU & ReLU, max pool]; 8, 16, 32 \\
Latent & 16 & $4\times4\times32$ \\
Decoder & FC 64, ReLU; & 3 $\times$ [t.conv $3\times3$, BN, \\
 & FC 256 (linear) & ReLU]; conv $3\times3$, sigmoid \\
Parameters & $3.5\times10^{4}$ & $2.1\times10^{4}$ \\
Mini-batch size & 32 & 64 \\ \hline
\end{tabular}
\end{table}

{\bf Raw IQ-samples.} We process the standardized \ac{PSD} vectors of Fig.~\ref{fig:iq_pipeline} through a \ac{FC} \ac{AE}. We configured the encoder to map the 256 frequency bins to 64 and then to 16 units, which form the latent representation, while the decoder mirrors the encoder. We configured the hidden layers to use \ac{ReLU} activations, while the output layer is linear, since the standardized bins can take both positive and negative values.

{\bf Symbols.} We processed the images of Fig.~\ref{fig:symbol_pipeline} through a convolutional \ac{AE}. We configured each of the three encoder stages with a $3\times3$ convolution, \ac{BN}, a \ac{ReLU}, and $2\times2$ max pooling, leading to a $4\times4\times32$ latent representation. As for the raw IQ, we configured the decoder to mirror the encoder, with stride-2 transposed convolutions and a convolution followed by a sigmoid, so as to map the output to the range of the input.

{\bf Training.} We trained both \acp{AE} with the objective of minimizing the \ac{MSE} with the Adam optimizer (learning rate $10^{-3}$) and the mini-batch sizes of Table~\ref{tab:ae}. We held out for validation the last 15\% of the training data of the transmitter observed during training, in time order. We evaluated the validation loss every 50 iterations, and we stopped training when it had not improved for six consecutive evaluations or after 6000 iterations, retaining the weights with the lowest validation loss. We used no \ac{OOD} example for model selection, in line with the open-set assumption. Moreover, we trained each configuration with three random seeds. We implemented all the models with the Deep Learning Toolbox provided by MATLAB, on an NVIDIA RTX 3090 \ac{GPU}.

\subsection{Training Strategies for the Symbols}
\label{sec:symbols_method}
Recall from Sect.~\ref{sec:preprocessing} that we generate the images per constellation point (symbol). Thus, the training of a transmitter with a given modulation provides one set of images for each symbol of that modulation. This provides several possible strategies for assigning such images to the models, i.e., how many \acp{AE} as presented in Sect.~\ref{sec:ae} we should use to describe a transmitter and how to process a test image through them.
This choice is central to our problem of cross-modulation \ac{RFFI} since, in general, the constellation points of the test modulation do not coincide with those observed during training.
Accordingly, in the following, we consider two strategies, which differ in the number of models associated with a transmitter: (i) a single \ac{AE} trained on the images of all the symbols of a modulation, and (ii) a pool of \acp{AE}, one for each symbol, where we process a test image through the four models whose symbols are the closest to its own. In both cases, we keep the cross-modulation \ac{RFFI} settings. Thus, we apply the models to a modulation different from the one used at training time, without re-training.

\begin{figure*}[t]
 \centering
 \includegraphics[width=\textwidth]{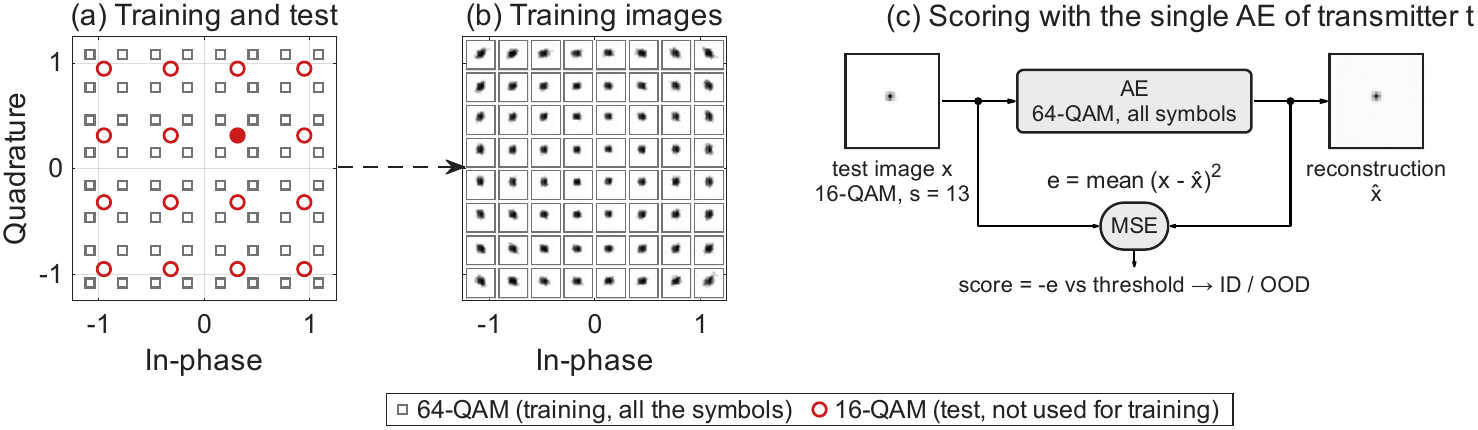}
 \caption{Single model applied to a modulation that was not observed during training, for the specific case of a transmitter trained with 64-QAM and a 16-QAM test image of symbol 13. (a) Constellation points of the trained modulation (grey squares) and of the test modulation (red circles) in the IQ plane, (b) training set of the autoencoder, (c) processing of the test image $x$.
 }
 \label{fig:single_method}
\end{figure*}

{\bf Single model.} We train one \ac{AE} for each transmitter and modulation on the images of all the symbols of that modulation, i.e., $140m$ images for the modulation order $m$. Therefore, the model learns the error vectors of the transmitter, independently of the constellation point they belong to. Accordingly, we describe a transmitter trained on all six modulations through six models, each trained with three seeds (Sect.~\ref{sec:ae}).
At test time, we directly feed each image to a trained \ac{AE}, regardless of the modulation order and symbol it represents. Thus, we provide the \ac{AE} with images generated from modulation orders that were not used during its training.
Figure~\ref{fig:single_method} illustrates this strategy for a transmitter trained with 64-QAM and tested with 16-QAM. Note that the training constellation and the test one do not share any point (Fig.~\ref{fig:single_method}a). We train the \ac{AE} on the images of all 64 symbols of the training modulation, whose error clouds change in shape and orientation with the position of the constellation point in the \ac{IQ} plane (Fig.~\ref{fig:single_method}b). Then, we process the test image through this \ac{AE} as is, and compare the \ac{MSE} between the image and its reconstruction to a threshold to accept or reject the claimed identity (Fig.~\ref{fig:single_method}c). Note that this strategy requires a single model to describe the images of all the constellation points of a modulation, although they do not necessarily share the same distribution of the error vectors.

\begin{figure*}[t]
 \centering
 \includegraphics[width=\textwidth]{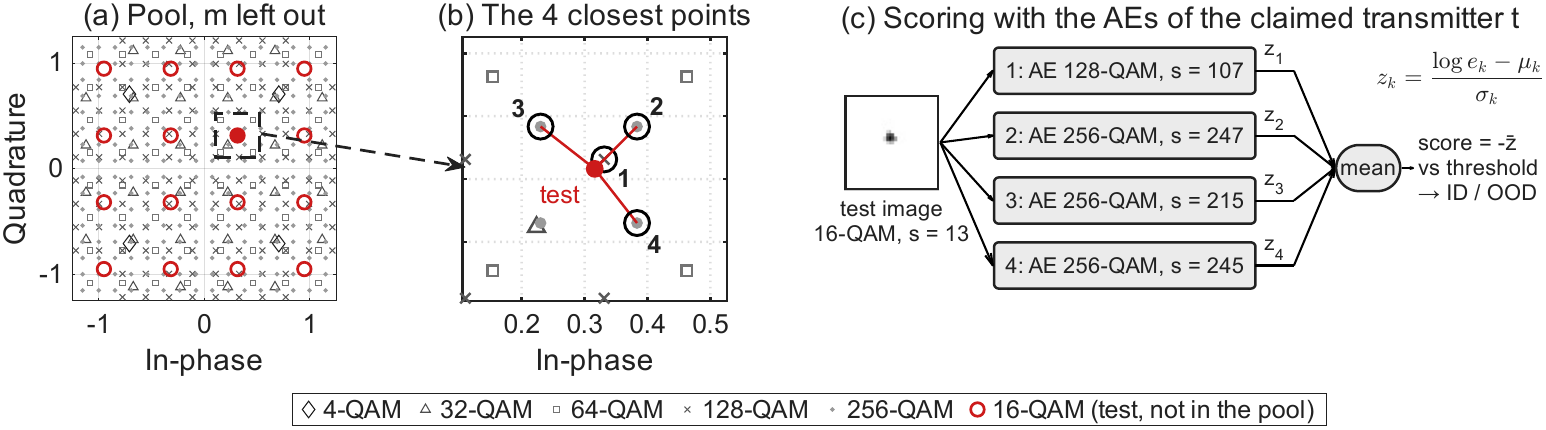}
 \caption{Pool of per-symbol models applied to a modulation that was not observed at training time, for the case of a 16-QAM test image of symbol 13. (a) Constellation points of the pool in the IQ plane (4-, 32-, 64-, 128- and 256-QAM, unit average power), (b) Zoom of the dashed square of (a), showing the four constellation points closest to the test symbol, numbered from the closest one, (c) processing of the test image.}
 \label{fig:pool_method}
\end{figure*}

{\bf Pool of per-symbol models.} In this strategy, we train one \ac{AE} for each transmitter, modulation, and symbol, on the 140 training images of that symbol only, leading to a pool of 500 \acp{AE} for a transmitter trained with all six modulations (one seed each, given their number). Therefore, each model describes the error vectors of a single constellation point, and it is more specific than the single model. On the other hand, the receiver has to select the models to apply to a test image. Since each image is generated from the (known) received symbols of one constellation point, we configured such a selection to rely on the modulation and the symbol of the image. If the receiver has observed during training the modulation, we process the image through the \ac{AE} of its own modulation and symbol; we refer to this strategy as {\em one-to-one}. Otherwise, we process the image through the four \acp{AE} of the pool whose constellation points are the closest to that of its symbol in the \ac{IQ} plane (see Appendix~\ref{sec:appendix} for the rationale of such an approach). We refer to this second strategy as {\em pool of per-symbol models}. We illustrate this procedure in Fig.~\ref{fig:pool_method} for a 16-QAM test image of symbol 13, when 16-QAM has not been observed during training. Here, the pool holds the constellation points of the other five modulations, each associated with the \ac{AE} of the claimed transmitter for that modulation and symbol (Fig.~\ref{fig:pool_method}a). Fig.~\ref{fig:pool_method}b shows the zoom of the four points closest to the test symbol, which are one 128-QAM point and three 256-QAM points. Finally, as shown in Fig.~\ref{fig:pool_method}c, we process the test image through the corresponding four \acp{AE}. Since the reconstruction errors of different \acp{AE} are not directly comparable, we standardize the logarithm of the error returned by each \ac{AE} through the mean and the standard deviation that the same \ac{AE} obtains on its validation images, and we then average the four standardized errors. Thus, we take the opposite of the average as the \ac{MSE} score of the image, and compare it with a threshold to accept or reject the claimed identity. The selection only requires the constellation of the new modulation.

We highlight that the two strategies discussed above represent opposite ends of a trade-off between complexity and specificity. The single model requires six \acp{AE} per transmitter and one evaluation per image, while the pool of \acp{AE} requires 500 \acp{AE} per transmitter, each trained on a small number of images, and four evaluations per image when the modulation was not observed during training. In return, in this latter strategy, we compare a test image only with the models of the constellation points that are the closest to its symbol, instead of using a model of the whole constellation. We investigate this trade-off experimentally in Sect.~\ref{sec:results}.

\section{Experimental Results}
\label{sec:results}

\subsection{RF Cable Scenario}
\label{sec:cable}
\begin{figure}[t]
 \centering
 \includegraphics[width=\columnwidth]{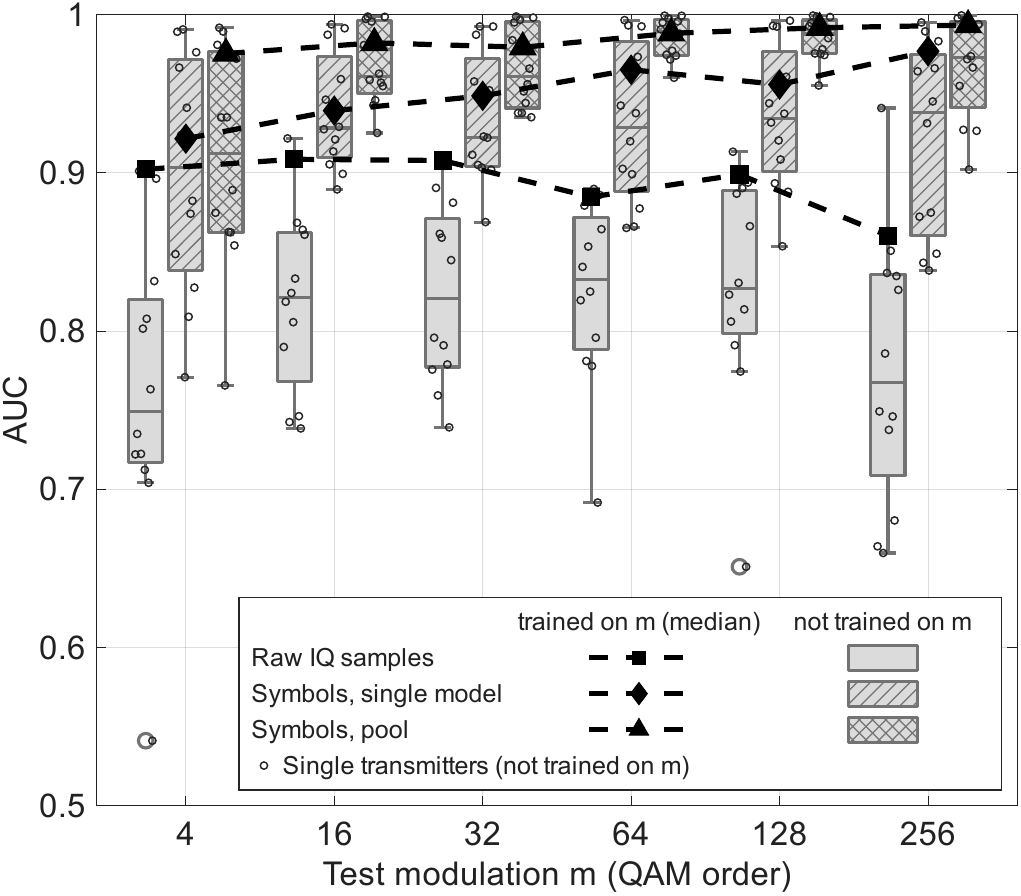}
 \caption{RF cable scenario: AUC as a function of the test modulation $m$ for the raw IQ samples (plain boxes, squares), the symbols with the single model (diagonal lines, diamonds) and the symbols with the pool of per-symbol models (crossed lines, triangles). Dashed curves: median AUC of the twelve transmitters when training on $m$. Boxes: AUC of the twelve transmitters (circles) when not training on $m$, averaged over the five other training modulations (raw IQ samples, single model) or with the pool without $m$.
 }
 \label{fig:cable}
\end{figure}

This section presents the baseline results of cross-modulation \ac{RFFI} in the \emph{RF cable} scenario, where the propagation channel does not come into play, for the two considered \ac{RFFI} approaches (Sect.~\ref{sec:symbols_method}). We evaluate the three configurations with the same one-vs-all protocol. We train on each of the twelve transmitters in turn as the authorized one, using only its training data; then, for every test modulation $m$, we distinguish the test samples of the reference transmitter (\ac{ID}) from those of the same modulation of the remaining eleven transmitters (\ac{OOD}). At testing time, for raw \ac{IQ} samples we consider a window of 2~ms of signal, while for the symbols we consider an image generated by considering 320 received symbols of the same constellation point. Each of these testing samples leads to one decision, i.e., accept (if the authorized transmitter is identified) or reject (if the model deems that this sample does not originate from the authorized transmitter).
We compute the \ac{AUC} for each authorized transmitter, in two cases: the matched one, in which the test modulation $m$ is observed at training time, and the one in which it is not. Figure~\ref{fig:cable} reports our results. For each test modulation $m$, the three grey box plots refer to the three configurations and differ by their fill pattern: from left to right, the raw \ac{IQ} samples (plain), the symbols with the single model (diagonal lines) and the symbols with the pool of per-symbol models (crossed lines). Each box plot summarizes the twelve \ac{AUC} values, one for each transmitter considered as authorized, achieved when $m$ was not observed at training time. The line inside the box is their median, while the edges of the box are their 25th and 75th percentiles, so the box spans the interquartile range and includes the central half of the transmitters. The whiskers extend to the lowest and to the highest values within 1.5 times the interquartile range from the edges of the box, while the values beyond the whiskers are outliers. The circles report the twelve values themselves. The three dashed black curves report the median over the twelve transmitters of the \ac{AUC} of the matched case, and they differ by their markers (squares, diamonds and triangles, respectively).
We discuss the results separately below for each approach.

{\bf Raw IQ samples.}
We train one \ac{AE} for each transmitter and modulation, and test it on all six modulations, i.e., for every test modulation, we have 500 test windows of the authorized transmitter and 5500 ($500\times11$) of the remaining ones. %
The median \ac{AUC} is approx. 0.90 in the classical \ac{RFFI} setting, i.e., when training and testing on the same modulation  (squares). In the cross-RFFI setting, the \ac{AUC} drops to approx. 0.82, and the loss grows with the distance between the modulation orders used at testing and training time: for example, the model trained on 4-QAM achieves 0.660 on 256-QAM. Moreover, the various heights of the boxes indicate that the performance changes considerably from one transmitter to another.

{\bf Symbols: single model.} We train one \ac{AE} for each transmitter and modulation on the images of all the symbols of that modulation, and test it on all six modulations, i.e., for every test modulation $m$, we have $60m$ test images of the authorized transmitter and $660m$ ($11\times60m$) of the remaining ones. %
When the training and the test modulations coincide (diamonds), the median \ac{AUC} grows with the modulation order, from 0.922 for 4-QAM to 0.977 for 256-QAM. When the modulation changes at testing time compared to the one used for training (boxes with diagonal lines), the \ac{AUC} varies slightly (variations in the second decimal digits), and the RFFI performance depends more on the transmitter than on the modulation. The loss concentrates on 4-QAM: every pair of different training and test modulations between 16-QAM and 256-QAM achieves at least 0.923, while the model trained on 4-QAM achieves AUC between 0.838 and 0.885 on the other modulations.

{\bf Symbols: pool of per-symbol models.} We consider the same test images as for the single model. We remove, in turn, each modulation $m$ from the pool and we use it for testing, as in the example of Fig.~\ref{fig:pool_method}. We refer to the matched case as the one-to-one case, in which the pool includes the test modulation. In the one-to-one case (triangles), the median \ac{AUC} grows from 0.976 (4-QAM) to 0.994 (256-QAM), and it exceeds that of the single model trained on the test modulation for every modulation. Thus, dedicating a model to each symbol improves both the performance and its uniformity across the transmitters. When the test modulation is removed from the pool (boxes with crossed lines), the median \ac{AUC} is at least 0.96 from 16-QAM on, within approx. 0.02 of the one-to-one case, while the 4-QAM is the exception, with 0.912. We use four models, as this value represents a reasonable trade-off: over the six modulations, the median \ac{AUC} is 0.967 with the closest model only, 0.975 with four models, and only 0.977 with eight.
\begin{finding}
    In the RF cable scenario, the fingerprint observed on the symbols largely persists across modulations, as opposed to the one observed on the raw \ac{IQ} samples. For the image-based approach working on symbols, the pool of per-symbol models (AUC=0.975) outperforms even the single model trained and tested on the same modulation (AUC=0.950).
\end{finding}

\subsection{Wireless LoS Scenario}
\label{sec:wireless}
\begin{figure}[t]
 \centering
 \includegraphics[width=\columnwidth]{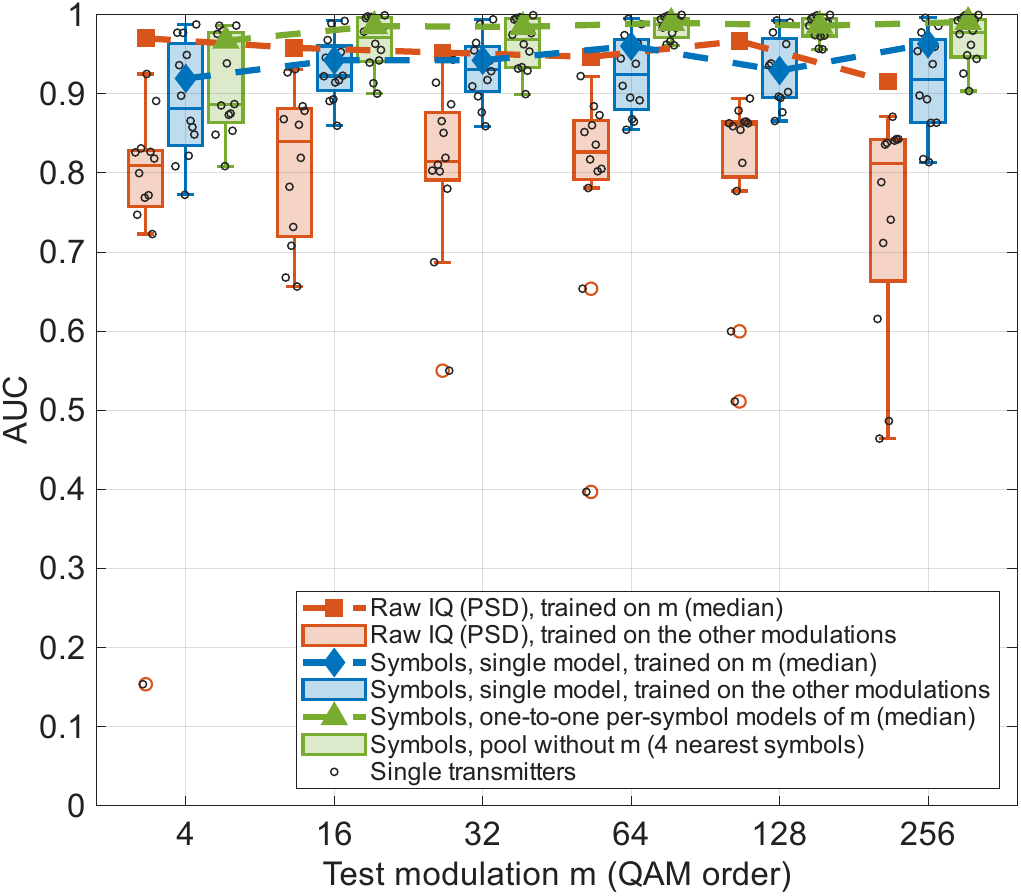}
 \caption{Wireless LoS scenario: AUC as a function of the test modulation $m$ for the raw IQ samples (orange), symbols with the single model (blue), and symbols with the pool of per-symbol models (green). Dashed curves: median AUC of the twelve transmitters when training on $m$. Boxes: AUC of the twelve transmitters (circles) when not training on $m$, averaged over the other five training modulations (raw IQ samples, single model) or with the pool without $m$.}
 \label{fig:wireless_joint}
\end{figure}

In this section, we consider the \emph{wireless} scenario, where the propagation channel comes into play. We use the same methodology as in Sect.~\ref{sec:cable}, and train the models from scratch on the wireless recordings, i.e., both the training and testing data are over-the-air data. We report in Fig.~\ref{fig:wireless_joint} the results for the three considered configurations using the same layout as of Fig.~\ref{fig:cable}.

{\bf Raw IQ samples.} As on the cable, the raw \ac{IQ} samples suffer the most from a change of modulation: when not training on the test modulation, the median \ac{AUC} is 0.823 (0.816 on the cable). However, the performance is less stable over the air: the standard deviation almost doubles, and the \ac{AUC} falls below 0.5 for four of the 72 pairs of transmitter and test modulation, i.e., worse than a random guess, while the minimum on the cable is 0.541. A plausible explanation is that the \ac{PSD}, estimated before any equalization, includes the frequency response of the channel, so a change of the channel leads the model to reject the authorized transmitter.

{\bf Symbols.} Conversely, the RFFI models working on symbols exhibit the same performance on over-the-air data as on cable data. For both training strategies, the median \ac{AUC} changes by less than 0.01 with respect to the \emph{RF cable} scenario, independently of training on the modulation considered for testing (Table~\ref{tab:summary}). The ranking of Sect.~\ref{sec:cable} is confirmed: the pool of per-symbol models is the best configuration and, when the test modulation is removed from it, it still outperforms the single model trained on the test modulation.

\begin{finding}
    Over a wireless \ac{LoS} link, the \ac{RFFI} models operating on symbols exhibit the same strong performance as observed on the cable: the median \ac{AUC} of both training strategies changes by less than 0.01.
    Conversely, the models operating on raw \ac{IQ} samples remain the least robust to a change of modulation (AUC=0.823) and become less stable: their \ac{AUC} falls below 0.5 for four of the 72 pairs of transmitter and test modulation.
\end{finding}

\begin{table}[t]
\caption{Comparison of the three configurations; the AUC is reported as median $\pm$ standard deviation over the twelve transmitters and the six test modulations $m$}
\label{tab:summary}
\centering
\footnotesize
\setlength{\tabcolsep}{4pt}
\begin{tabular}{|l|c|c|c|}
\hline
 & \textbf{Raw IQ} & \textbf{Symbols,} & \textbf{Symbols,} \\
 & \textbf{samples} & \textbf{single model} & \textbf{pool} \\ \hline
Signal per decision & 2~ms & 320 symbols & 320 symbols \\
Time to collect it & 2~ms & 7~ms--0.46~s & 7~ms--0.46~s \\
AEs per transmitter & 6 & 6 & 500 \\ \hline
\multicolumn{4}{|l|}{\textbf{RF cable}} \\
AUC, training on $m$ & $0.897\pm0.047$ & $0.950\pm0.034$ & $0.987\pm0.012$ \\
AUC, not training on $m$ & $0.816\pm0.075$ & $0.928\pm0.053$ & $0.975\pm0.043$ \\ \hline
\multicolumn{4}{|l|}{\textbf{Wireless LoS}} \\
AUC, training on $m$ & $0.951\pm0.156$ & $0.945\pm0.037$ & $0.985\pm0.015$ \\
AUC, not training on $m$ & $0.823\pm0.136$ & $0.926\pm0.054$ & $0.976\pm0.043$ \\ \hline
\end{tabular}
\end{table}

{\bf Summary.} Table~\ref{tab:summary} summarizes the results of both scenarios. In both, the pool of per-symbol models achieves the highest median \ac{AUC} and the lowest standard deviation, and the models operating on symbols outperform the ones working on raw \ac{IQ} samples when the test modulation was not observed during training. The three configurations rely on different amounts of signal for a decision and on a different number of models. Finally, when not observed during training, 4-QAM is the most critical modulation in all configurations and scenarios, possibly because it has the lowest number of points in the IQ plane, all its points have the same amplitude, and, given the fixed peak level of Sect.~\ref{sec:setup}, it is transmitted with the highest average power.

\subsection{Cross-Scenario Analysis}
\label{sec:crossscenario}
\begin{figure}[t]
 \centering
 \includegraphics[width=\columnwidth]{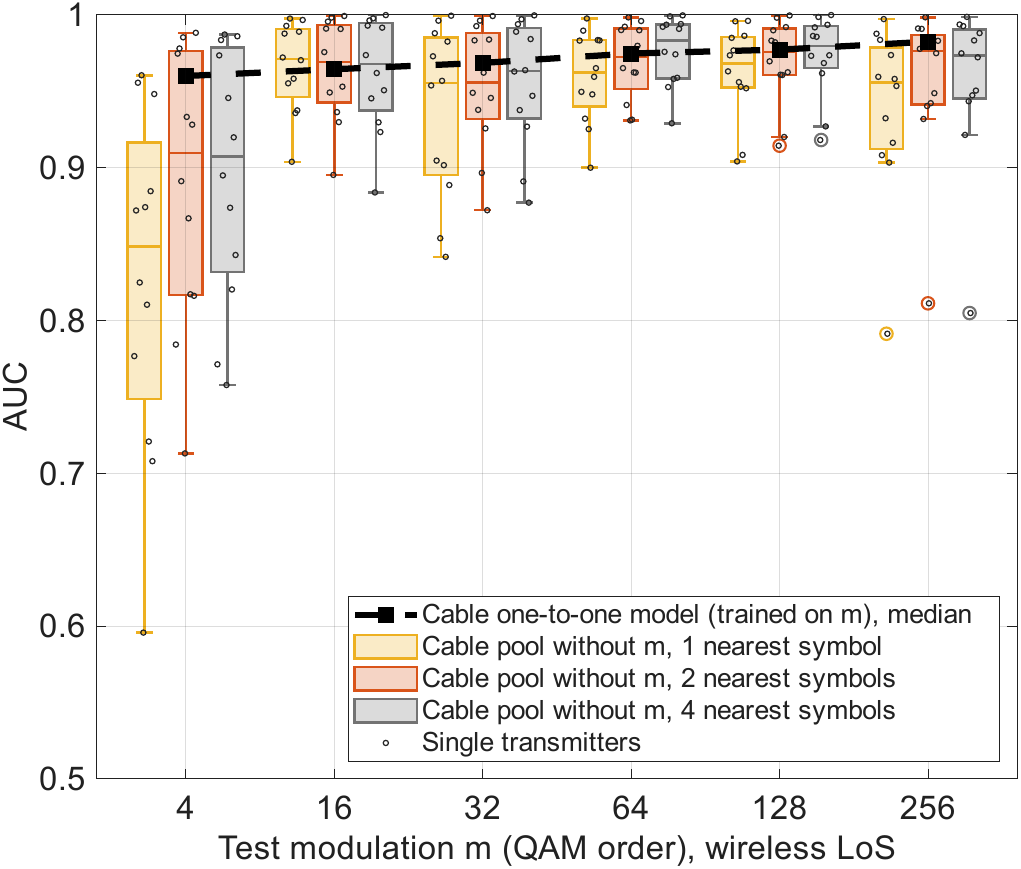}
 \caption{Cross-scenario: per-symbol autoencoders trained on the RF cable and tested on the wireless LoS recordings. Dashed curve: median AUC of the twelve transmitters with the one-to-one model of the test modulation $m$. Boxes: AUC of the twelve transmitters (circles) with the pool without $m$, scoring with the 1, 2 or 4 closest symbols.}
 \label{fig:crossscenario_pool}
\end{figure}

In the previous subsections, we trained and tested the transmitters with data collected in the same scenario. Here, we consider the pools of per-symbol \acp{AE} trained in the \emph{RF cable} scenario, and we test them on the images generated from the data of the \emph{wireless} scenario, without any re-training. Figure~\ref{fig:crossscenario_pool} reports the results with the same layout of Fig.~\ref{fig:cable}: the three boxes of each test modulation refer to the pool without $m$ when testing on an image using the models of the one, two or four closest symbols. When training and testing on the same modulation (dashed curve), the median \ac{AUC} is 0.973, against the \ac{AUC} value 0.985 obtained for the models trained on-over-the-air measurements (Table~\ref{tab:summary}), i.e., the change of link costs only approx. 0.01 in terms of AUC. When the test modulation is not included in the training set, the pool with the four closest symbols achieves AUC=0.973, against AUC=0.976 for the pool trained on over-the-air data. Thus, it still outperforms the single model trained and tested on over-the-air data including the test modulation itself (AUC=0.945). Finally, the number of closest symbols considered for training the model matters mostly for 4-QAM, whose median \ac{AUC} grows from 0.848 with one symbol to about 0.91 with two or four.

\begin{finding}
    The fingerprint observed on the symbols is preserved across scenarios: the pool of per-symbol models trained on cable data achieves a median \ac{AUC} of 0.973 when tested on over-the-air data on a modulation that was not part of the training set, against AUC=0.976 when trained on over-the-air data.
\end{finding}

\section{Conclusions}
\label{sec:conclusions}
In this paper, we analyzed cross-modulation \ac{RFFI} in an open-set setting, with twelve nominally identical transmitters, six modulation orders from 4-QAM to 256-QAM, and two links. Our results show that the type of data used for RFFI is a key choice: the fingerprint observed by models operating on the symbols is largely preserved when the modulation changes, as opposed to the one observed on the raw \ac{IQ} samples, which is also less stable over the air. Moreover, through a pool of per-symbol models, we can identify a transmitter even when using a modulation that was never observed at training time, with a median \ac{AUC} of approx. 0.975, even when training on the cable and testing over the air. Future work includes tests with links without \ac{LoS}, mobility, and the stability of the fingerprint over time.

\bibliographystyle{ACM-Reference-Format}
\bibliography{references}

@inproceedings{suski2008spectral,
  author    = {Suski, II, William C. and Temple, Michael A. and Mendenhall, Michael J. and Mills, Robert F.},
  title     = {Using Spectral Fingerprints to Improve Wireless Network Security},
  booktitle = {Proc. IEEE Global Telecommunications Conference (GLOBECOM)},
  year      = {2008},
  pages     = {1--5}
}

@inproceedings{oligeri2026wisec,
  title={{Beyond Static Signatures: Statistical Analysis of Radio Fingerprint Mutations}},
  author={Oligeri, Gabriele and Sciancalepore, Savio},
  booktitle={Proceedings of the 19th ACM Conference on Security and Privacy in Wireless and Mobile Networks},
  pages={239--249},
  year={2026}
}

@inproceedings{kennedy2008steady,
  author    = {Kennedy, Irwin O. and Scanlon, Patricia and Mullany, Francis J. and Buddhikot, Milind M. and Nolan, Keith E. and Rondeau, Thomas W.},
  title     = {Radio Transmitter Fingerprinting: A Steady State Frequency Domain Approach},
  booktitle = {Proc. IEEE Vehicular Technology Conference (VTC-Fall)},
  year      = {2008},
  pages     = {1--5}
}

@inproceedings{shen2021lora,
  author    = {Shen, Guanxiong and Zhang, Junqing and Marshall, Alan and Peng, Linning and Wang, Xianbin},
  title     = {Radio Frequency Fingerprint Identification for {LoRa} Using Spectrogram and {CNN}},
  booktitle = {Proc. IEEE Conference on Computer Communications (INFOCOM)},
  year      = {2021},
  doi       = {10.1109/INFOCOM42981.2021.9488793}
}

@article{welch1967,
  author  = {Welch, Peter D.},
  title   = {The Use of Fast {Fourier} Transform for the Estimation of Power Spectra: A Method Based on Time Averaging Over Short, Modified Periodograms},
  journal = {IEEE Transactions on Audio and Electroacoustics},
  volume  = {15},
  number  = {2},
  pages   = {70--73},
  year    = {1967},
  doi     = {10.1109/TAU.1967.1161901}
}

@inproceedings{brik2008paradis,
  author    = {Brik, Vladimir and Banerjee, Suman and Gruteser, Marco and Oh, Sangho},
  title     = {Wireless Device Identification with Radiometric Signatures},
  booktitle = {Proc. ACM International Conference on Mobile Computing and Networking (MobiCom)},
  year      = {2008},
  pages     = {116--127}
}

@article{peng2020dctf,
  author  = {Peng, Linning and Zhang, Junqing and Liu, Ming and Hu, Aiqun},
  title   = {Deep Learning Based {RF} Fingerprint Identification Using Differential Constellation Trace Figure},
  journal = {IEEE Transactions on Vehicular Technology},
  volume  = {69},
  number  = {1},
  pages   = {1091--1095},
  year    = {2020},
  doi     = {10.1109/TVT.2019.2950670}
}

@article{oligeri2023pastai,
  author  = {Oligeri, Gabriele and Sciancalepore, Savio and Raponi, Simone and Di Pietro, Roberto},
  title   = {{PAST-AI}: Physical-Layer Authentication of Satellite Transmitters via Deep Learning},
  journal = {IEEE Transactions on Information Forensics and Security},
  volume  = {18},
  pages   = {274--289},
  year    = {2023},
  doi     = {10.1109/TIFS.2022.3219287}
}

@inproceedings{alhazbi2023dat,
  author    = {Alhazbi, Saeif and Sciancalepore, Savio and Oligeri, Gabriele},
  title     = {The Day-After-Tomorrow: On the Performance of Radio Fingerprinting over Time},
  booktitle = {Proc. Annual Computer Security Applications Conference (ACSAC)},
  year      = {2023},
  pages     = {439--450},
  doi       = {10.1145/3627106.3627192}
}

@article{papangelo2024adversarial,
  author  = {Papangelo, Lorenzo and Pistilli, Maurizio and Sciancalepore, Savio and Oligeri, Gabriele and Piro, Giuseppe and Boggia, Gennaro},
  title   = {Adversarial Machine Learning for Image-Based Radio Frequency Fingerprinting: Attacks and Defenses},
  journal = {IEEE Communications Magazine},
  volume  = {62},
  number  = {11},
  pages   = {108--113},
  year    = {2024},
  doi     = {10.1109/MCOM.001.2300464}
}

@inproceedings{irfan2025jamming,
  author    = {Irfan, Muhammad and Sciancalepore, Savio and Oligeri, Gabriele},
  title     = {Preventing Radio Fingerprinting through Low-Power Jamming},
  booktitle = {Proc. ACM Asia Conference on Computer and Communications Security (ASIA CCS)},
  year      = {2025},
  pages     = {1114--1126},
  doi       = {10.1145/3708821.3733862}
}

@inproceedings{oligeri2026hideprint,
  author    = {Oligeri, Gabriele and Sciancalepore, Savio},
  title     = {{HidePrint}: Protecting Device Anonymity by Obscuring Radio Fingerprints},
  booktitle = {Proc. ACM Asia Conference on Computer and Communications Security (ASIA CCS)},
  year      = {2026},
  pages     = {250--262},
  doi       = {10.1145/3779208.3785266}
}

@article{sciancalepore2024jamming,
  author  = {Sciancalepore, Savio and Kusters, Fabrice and Abdelhadi, Nada Khaled and Oligeri, Gabriele},
  title   = {Jamming Detection in Low-{BER} Mobile Indoor Scenarios via Deep Learning},
  journal = {IEEE Internet of Things Journal},
  volume  = {11},
  number  = {8},
  pages   = {14682--14697},
  year    = {2024},
  doi     = {10.1109/JIOT.2023.3343611}
}

@inproceedings{alshawabka2020exposing,
  author    = {Al-Shawabka, Amani and Restuccia, Francesco and D'Oro, Salvatore and Jian, Tong and Costa Rendon, Bruno and Soltani, Nasim and Dy, Jennifer and Ioannidis, Stratis and Chowdhury, Kaushik and Melodia, Tommaso},
  title     = {Exposing the Fingerprint: Dissecting the Impact of the Wireless Channel on Radio Fingerprinting},
  booktitle = {Proc. IEEE Conference on Computer Communications (INFOCOM)},
  year      = {2020},
  pages     = {646--655},
  doi       = {10.1109/INFOCOM41043.2020.9155259}
}

@inproceedings{tyler2022intentional,
  author    = {Tyler, Joshua H. and Fadul, Mohamed K. M. and Reising, Donald R. and Liang, Yu},
  title     = {Assessing the Presence of Intentional Waveform Structure In Preamble-based {SEI}},
  booktitle = {Proc. IEEE Global Communications Conference (GLOBECOM)},
  year      = {2022},
  pages     = {4129--4135},
  doi       = {10.1109/GLOBECOM48099.2022.10001025}
}

@misc{wong2018iqimbalance,
  author        = {Wong, Lauren J. and Headley, William C. and Michaels, Alan J.},
  title         = {Emitter Identification Using {CNN} {IQ} Imbalance Estimators},
  year          = {2018},
  eprint        = {1808.02369},
  archivePrefix = {arXiv},
  note          = {arXiv:1808.02369}
}

@inproceedings{agadakos2020chameleons,
  author    = {Agadakos, Ioannis and Agadakos, Nikolaos and Polakis, Jason and Amer, Mohamed R.},
  title     = {Chameleons' Oblivion: Complex-Valued Deep Neural Networks for Protocol-Agnostic {RF} Device Fingerprinting},
  booktitle = {Proc. IEEE European Symposium on Security and Privacy (EuroS\&P)},
  year      = {2020},
  pages     = {322--338},
  doi       = {10.1109/EuroSP48549.2020.00028}
}

@article{huang2024dtp,
  author  = {Huang, Da and Al-Hourani, Akram and Sithamparanathan, Kandeepan and Rowe, Wayne S. T.},
  title   = {Deep Learning Methods for {IoT} Device Authentication Using Symbols Density Trace Plot},
  journal = {IEEE Internet of Things Journal},
  volume  = {11},
  number  = {10},
  pages   = {18167--18179},
  year    = {2024},
  doi     = {10.1109/JIOT.2024.3361892}
}

@article{zhang2023variable,
  author  = {Zhang, Xinliang and Li, Tianyun and Gong, Pei and Zha, Xiong and Liu, Renwei},
  title   = {Variable-Modulation Specific Emitter Identification With Domain Adaptation},
  journal = {IEEE Transactions on Information Forensics and Security},
  volume  = {18},
  pages   = {380--395},
  year    = {2023},
  doi     = {10.1109/TIFS.2022.3223794}
}

@inproceedings{yin2023fewshot,
  author    = {Yin, Lijun and Fu, Xue and Shi, Shengnan and Liu, Pengfei and Lin, Yun and Wang, Yu and Sari, Hikmet},
  title     = {Few-Shot Domain Adaption-Based Specific Emitter Identification Under Varying Modulation},
  booktitle = {Proc. IEEE International Conference on Communication Technology (ICCT)},
  year      = {2023},
  pages     = {1439--1443},
  doi       = {10.1109/ICCT59356.2023.10419733}
}

@article{zhang2025mdd,
  author  = {Zhang, Yezhuo and Zhou, Zinan and Cao, Yichao and Li, Guangyu and Li, Xuanpeng},
  title   = {Improved Specific Emitter Identification Based on Margin Disparity Discrepancy in Varying Modulation Scenarios},
  journal = {IEEE Signal Processing Letters},
  volume  = {32},
  pages   = {3375--3379},
  year    = {2025},
  doi     = {10.1109/LSP.2025.3597100}
}

@inproceedings{he2025crossmod,
  author    = {He, Yazhe and Zhang, Qin},
  title     = {Cross-Modulation Specific Emitter Identification Based on Source-Free Domain Adaptation},
  booktitle = {Proc. International Conference on Computer and Communication Systems (ICCCS)},
  year      = {2025},
  pages     = {458--463},
  doi       = {10.1109/ICCCS65393.2025.11069648}
}

@article{shi2026diva,
  author  = {Shi, Feng and Zhang, Xixi and Tang, Tiantian and Wang, Yu and Ohtsuki, Tomoaki and Gui, Guan},
  title   = {Robust Specific Emitter Identification Across Modulation Domains via Domain-Invariant Variational Autoencoding},
  journal = {IEEE Transactions on Information Forensics and Security},
  volume  = {21},
  pages   = {5880--5891},
  year    = {2026},
  doi     = {10.1109/TIFS.2026.3705305}
}

@article{wang2026rsfe,
  author  = {Wang, Dengxi and Huang, Keju and Sun, Jiaxing and Liu, Hui and Yang, Junan},
  title   = {Reconstructed-Signal-Assisted Domain Generalization for Variable-Modulation Specific Emitter Identification},
  journal = {IEEE Internet of Things Journal},
  year    = {2026},
  note    = {Early access},
  doi     = {10.1109/JIOT.2026.3747057}
}

@article{hanna2021openset,
  author  = {Hanna, Samer and Karunaratne, Samurdhi and Cabric, Danijela},
  title   = {Open Set Wireless Transmitter Authorization: Deep Learning Approaches and Dataset Considerations},
  journal = {IEEE Transactions on Cognitive Communications and Networking},
  volume  = {7},
  number  = {1},
  pages   = {59--72},
  year    = {2021},
  doi     = {10.1109/TCCN.2020.3043332}
}

@inproceedings{smailes2023watch,
  author    = {Smailes, Joshua and K{\"o}hler, Sebastian and Birnbach, Simon and Strohmeier, Martin and Martinovic, Ivan},
  title     = {Watch This Space: Securing Satellite Communication through Resilient Transmitter Fingerprinting},
  booktitle = {Proc. ACM SIGSAC Conference on Computer and Communications Security (CCS)},
  year      = {2023},
  pages     = {608--621},
  doi       = {10.1145/3576915.3623135}
}

@article{wang2026diame,
  author  = {Wang, Rui and Chen, Si and Cai, Zhenxin and Wang, Qin and Liu, Chengcheng and Lin, Yun and Gui, Guan},
  title   = {Bridging Modulation Gaps: Similarity-Aware Domain-Invariant Learning for Robust Radio Frequency Fingerprint Identification},
  journal = {IEEE Internet of Things Journal},
  volume  = {13},
  number  = {10},
  pages   = {22183--22193},
  year    = {2026},
  doi     = {10.1109/JIOT.2026.3668282}
}

@inproceedings{chen2024agnostic,
  author    = {Chen, Yanghong and Xu, Xiaodong},
  title     = {A Modulation Agnostic Specific Emitter Identification Scheme Based on Expert Knowledge Informed Deep Learning},
  booktitle = {Proc. International Conference on Communications, Circuits and Systems (ICCCAS)},
  year      = {2024},
  pages     = {321--326},
  doi       = {10.1109/ICCCAS62034.2024.10652841}
}

@misc{chen2025disentangled,
  author       = {Chen, Yanghong and Xu, Xiaodong and Qin, Xiaowei},
  title        = {Disentangled Generative Model Based Specific Emitter Identification Scheme with Visual Interpretation},
  howpublished = {SSRN preprint 5226582},
  year         = {2025},
  url          = {https://ssrn.com/abstract=5226582}
}

@inproceedings{saifaldawla2023ngso,
  author    = {Saifaldawla, Almoatssimbillah and Ortiz-Gomez, Flor G. and Lagunas, Eva and Daoud, Saed and Chatzinotas, Symeon},
  title     = {{NGSO}-To-{GSO} Satellite Interference Detection Based on Autoencoder},
  booktitle = {Proc. IEEE International Symposium on Personal, Indoor and Mobile Radio Communications (PIMRC)},
  year      = {2023},
  pages     = {1--7},
  doi       = {10.1109/PIMRC56721.2023.10293839}
}

@article{saifaldawla2024genai,
  author  = {Saifaldawla, Almoatssimbillah and Ortiz, Flor and Lagunas, Eva and Adam, Abuzar B. M. and Chatzinotas, Symeon},
  title   = {{GenAI}-Based Models for {NGSO} Satellites Interference Detection},
  journal = {IEEE Transactions on Machine Learning in Communications and Networking},
  volume  = {2},
  pages   = {904--924},
  year    = {2024},
  doi     = {10.1109/TMLCN.2024.3418933}
}

@article{rajendran2019saife,
  author  = {Rajendran, Sreeraj and Meert, Wannes and Lenders, Vincent and Pollin, Sofie},
  title   = {Unsupervised Wireless Spectrum Anomaly Detection With Interpretable Features},
  journal = {IEEE Transactions on Cognitive Communications and Networking},
  volume  = {5},
  number  = {3},
  pages   = {637--647},
  year    = {2019},
  doi     = {10.1109/TCCN.2019.2911524}
}

@article{shi2026psd2san,
  author  = {Shi, Wenqiang and Lei, Yingke and Teng, Fei and Jin, Hu and Wang, Jin and Pan, Bisheng and Yan, Xu and Pei, Ke},
  title   = {{PSD$_2$SAN}: An Unsupervised Specific Emitter Cross-Domain Identification Framework Under Class Inequality Conditions},
  journal = {IEEE Transactions on Cognitive Communications and Networking},
  volume  = {12},
  pages   = {10265--10280},
  year    = {2026},
  doi     = {10.1109/TCCN.2026.3706517}
}

@inproceedings{liu2008nonlinear,
  author    = {Liu, Ming-Wei and Doherty, John F.},
  title     = {Specific Emitter Identification using Nonlinear Device Estimation},
  booktitle = {Proc. IEEE Sarnoff Symposium},
  year      = {2008},
  pages     = {1--5},
  doi       = {10.1109/SARNOF.2008.4520119}
}

@inproceedings{xie2025regrowth,
  author    = {Xie, Lingnan and Peng, Linning and Zhang, Junqing},
  title     = {Towards Robust {RF} Fingerprint Identification Using Spectral Regrowth and Carrier Frequency Offset},
  booktitle = {Proc. IEEE Conference on Computer Communications (INFOCOM)},
  year      = {2025},
  pages     = {1--10},
  doi       = {10.1109/INFOCOM55648.2025.11044651}
}

@inproceedings{elmaghbub2020widescan,
  author    = {Elmaghbub, Abdurrahman and Hamdaoui, Bechir and Natarajan, Arun},
  title     = {{WideScan}: Exploiting Out-of-Band Distortion for Device Classification Using Deep Learning},
  booktitle = {Proc. IEEE Global Communications Conference (GLOBECOM)},
  year      = {2020},
  pages     = {1--6},
  doi       = {10.1109/GLOBECOM42002.2020.9348138}
}

@article{cai2022uav,
  author  = {Cai, Zhuoran and Liu, Zhiyuan and Kou, Liang},
  title   = {Reliable {UAV} Monitoring System Using Deep Learning Approaches},
  journal = {IEEE Transactions on Reliability},
  volume  = {71},
  number  = {2},
  pages   = {973--983},
  year    = {2022},
  doi     = {10.1109/TR.2021.3119068}
}

@article{jagannath2022survey,
  author  = {Jagannath, Anu and Jagannath, Jithin and Kumar, Prem Sagar Pattanshetty Vasanth},
  title   = {A Comprehensive Survey on Radio Frequency ({RF}) Fingerprinting: Traditional Approaches, Deep Learning, and Open Challenges},
  journal = {Computer Networks},
  volume  = {219},
  pages   = {109455},
  year    = {2022},
  doi     = {10.1016/j.comnet.2022.109455}
}

@misc{scholl2026dsp,
  author        = {Scholl, Stefan},
  title         = {{RF} Fingerprinting of {DSP} Implementations using Machine Learning: An Over-the-Air Case Study},
  year          = {2026},
  eprint        = {2610.15494},
  archivePrefix = {arXiv},
  note          = {arXiv:2610.15494}
}

@article{goldsmith1997variable,
  author  = {Goldsmith, Andrea J. and Chua, Soon-Ghee},
  title   = {Variable-Rate Variable-Power {MQAM} for Fading Channels},
  journal = {IEEE Transactions on Communications},
  volume  = {45},
  number  = {10},
  pages   = {1218--1230},
  year    = {1997},
  doi     = {10.1109/26.634685}
}

@article{svensson2007adaptive,
  author  = {Svensson, Arne},
  title   = {An Introduction to Adaptive {QAM} Modulation Schemes for Known and Predicted Channels},
  journal = {Proceedings of the IEEE},
  volume  = {95},
  number  = {12},
  pages   = {2322--2336},
  year    = {2007},
  doi     = {10.1109/JPROC.2007.904442}
}

@ARTICLE{5g_link_adapt,
  author={Pocovi, Guillermo and Pedersen, Klaus I. and Mogensen, Preben},
  journal={IEEE Access},
  title={Joint Link Adaptation and Scheduling for 5G Ultra-Reliable Low-Latency Communications},
  year={2018},
  volume={6},
  number={},
  pages={28912-28922},
  doi={10.1109/ACCESS.2018.2838585}}

@INPROCEEDINGS{80211n_adaptive,
  author={Peng, Fei and Zhang, Jinyun and Ryan, William E.},
  booktitle={2007 IEEE Wireless Communications and Networking Conference},
  title={Adaptive Modulation and Coding for IEEE 802.11n},
  year={2007},
  volume={},
  number={},
  pages={656-661},
  doi={10.1109/WCNC.2007.126}}

\appendix
\section{Position of the Symbols in the IQ Plane}
\label{sec:appendix}

\begin{figure}[!b]
 \centering
 \includegraphics[width=\columnwidth]{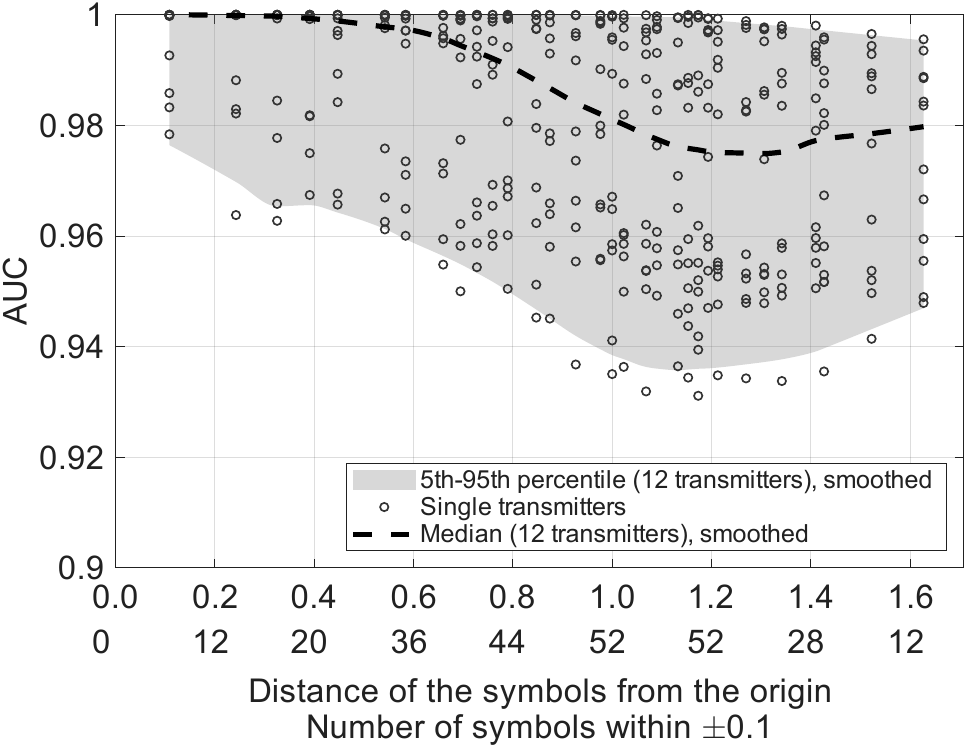}
 \caption{RF cable scenario, 256-QAM: AUC as a function of the distance of the symbols from the origin of the IQ plane, with one autoencoder per transmitter and ring of symbols at the same distance.}
 \Description{Scatter plot of the AUC of the twelve transmitters versus the distance of the symbols from the origin, from 0 to 1.6, with a grey area for the 5th to 95th percentiles and a dashed curve for the median. The median is close to 1 up to a distance of about 0.5, decreases to about 0.975 at 1.3 and slightly increases at the end. A second row of labels gives the number of symbols around each tick, from 0 to 52.}
 \label{fig:app_distance}
\end{figure}

\begin{figure}[!b]
 \centering
 \includegraphics[width=\columnwidth]{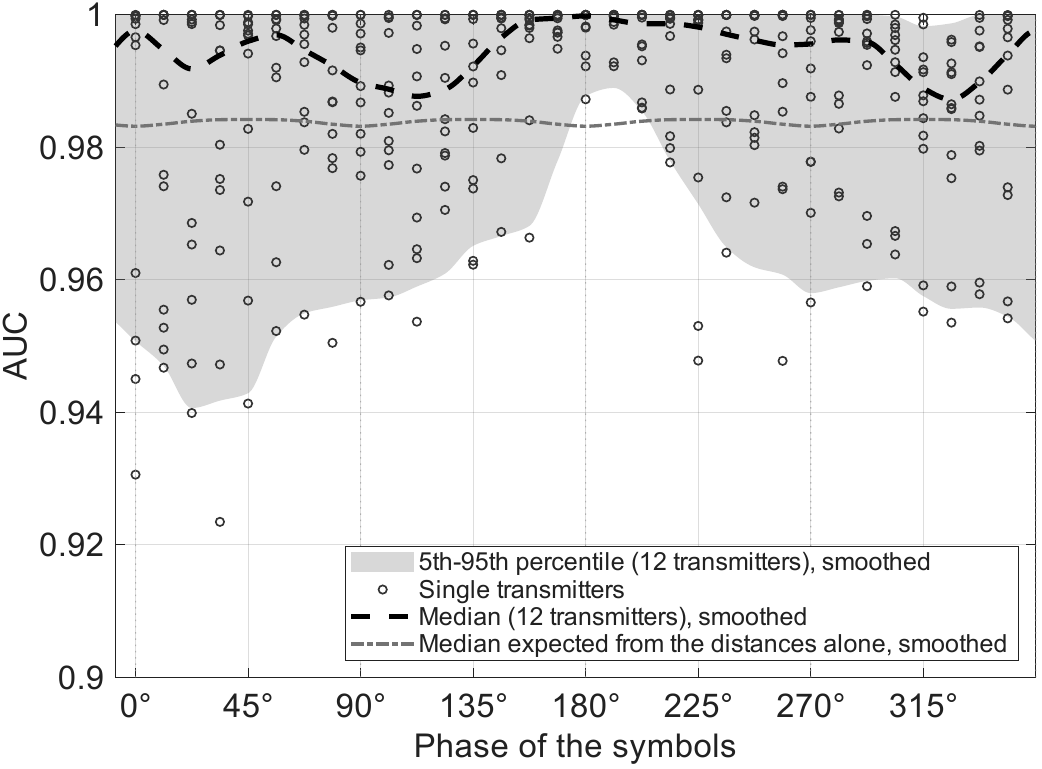}
 \caption{RF cable scenario, 256-QAM: AUC as a function of the phase of the symbols in the IQ plane, with one autoencoder per transmitter and sector of 11.25\textdegree.}
 \Description{Scatter plot of the AUC of the twelve transmitters versus the phase of the symbols, from 0 to 360 degrees, with a grey area for the 5th to 95th percentiles, a dashed curve for the median, between 0.987 and 1 at all phases, and a flat dash-dotted curve at about 0.983 for the median expected from the distances alone.}
 \label{fig:app_phase}
\end{figure}

When testing on a modulation not part of the training set, we use the {\em pool of per-symbol models} strategy (Sect.~\ref{sec:symbols_method}). We designed such a strategy to select the models to apply for \ac{RFFI} by considering the proximity of the constellation points in the \ac{IQ} plane.
In this appendix, we provide experimental support for this criterion, by showing that the \ac{AUC} achieved by such a model depends on the position of the symbols in the \ac{IQ} plane. We consider the \emph{RF cable} scenario and the 256-QAM, i.e., the densest constellation, normalized to unit average power. We group the 256 symbols in two ways: (i) by their distance from the origin, into 32 rings, each including 4 to 16 symbols at the same distance, and (ii) by their phase, into 32 sectors of 11.25\textdegree, each including 6 to 12 symbols. For every group and transmitter, we train one \ac{AE} on the training images of the symbols of the group, with the settings of Sect.~\ref{sec:ae} and three seeds, and we evaluate it with the one-vs-all protocol of Sect.~\ref{sec:cable} on the test images of the same symbols. Figures~\ref{fig:app_distance} and~\ref{fig:app_phase} report the \ac{AUC} of each transmitter, averaged over the seeds (circles), the range between the 5th and the 95th percentiles over the twelve transmitters (grey area), and their median (dashed curve); the edges of the area and the median are smoothed with a local regression over the distance or phase, respectively. In Fig.~\ref{fig:app_distance}, the second row of the horizontal axis reports the number of symbols around each tick.

{\bf Distance.} The \ac{AUC} changes gradually with the distance of the symbols from the origin (Fig.~\ref{fig:app_distance}). The median is close to 1 for the innermost rings, it decreases to about 0.975 at a distance of about 1.3 units, and it slightly increases toward the corners of the constellation, with no abrupt change between neighboring rings. Therefore, symbols at similar distances from the origin, as are those that are close in the \ac{IQ} plane, achieve a similar \ac{AUC}, which is consistent with the selection of the closest constellation points used in the paper for the pool of per-symbol models.

{\bf Phase.} Conversely, the phase of the symbols does not  affect significantly the \ac{AUC} (Fig.~\ref{fig:app_phase}). The median AUC remains between 0.987 and 1 for all the phases, and it is never lower than the median expected from the distances of the symbols of each sector alone (dash-dotted curve), i.e., the average over the symbols of the sector of the median \ac{AUC} of their rings. Thus, the phase penalizes no group of symbols, and only the spread of the weakest transmitters changes with it. Accordingly, the selection of the symbols in the pool of per symbols model relies on their proximity in the \ac{IQ} plane, with no specific constraint on their phase.

\end{document}